\documentclass[twocolumn,aps,amsmath,amssymb]{revtex4-1}
\usepackage{mathtools,siunitx, xcolor}
\usepackage[normalem]{ulem}
\usepackage[colorinlistoftodos]{todonotes}
\usepackage{soul}

\DeclarePairedDelimiter\bra{\langle}{\rvert}
\DeclarePairedDelimiter\ket{\lvert}{\rangle}

\begin{document}

\title{Effects of the measurement power on states discrimination and dynamics in a circuit-QED experiment}

\author{L. Tosi$^{1,2}$, I. Lobato$^{3}$, M. F. Goffman$^{1}$, C. Metzger$^{1}$, C. Urbina$^{1}$, and H. Pothier$^{1}$}
\email[]{hugues.pothier@cea.fr}
\affiliation{
$^{1}$Quantronics group, Service de Physique de l'\'Etat Condens\'e  (CNRS, UMR\ 3680),\\IRAMIS, CEA-Saclay, Universit\'e Paris-Saclay, 91191 Gif-sur-Yvette, France\\
$^{2}$Devices and Sensors Group; $^{3}$Balseiro Institute,\\ Centro At\'omico Bariloche, CNEA, CONICET, Argentina.}

\date{\today}

\begin{abstract}
We explore the effects of driving a cavity at a large photon number in a circuit-QED experiment where the ``matter-like'' part corresponds to an unique Andreev level in a superconducting weak link. The three many-body states of the weak link, corresponding to the occupation of the Andreev level by 0, 1 or 2 quasiparticles, lead to different cavity frequency shifts. We show how the non-linearity inherited by the cavity from its coupling to the weak link affects the state discrimination and the photon number calibration. Both effects require treating the evolution of the driven system beyond the dispersive limit. In addition, we observe how transition rates between the circuit states (quantum and parity jumps) are affected by the microwave power, and compare the measurements with a theory accounting for the ``dressing'' of the Andreev states by the cavity.

\end{abstract}
\maketitle

\section{Introduction} 
Circuit-QED (cQED) describes the coupling between a quantized mode of the electromagnetic field in a microwave cavity and the atomic-like energy levels of a quantum circuit. In practice, the quantum state of the circuit  is very efficiently accessed through its effect on the coupled cavity, thus making cQED a method of choice for probing superconducting devices \cite{Blais2021}. In experiments aiming at the quantum manipulation of the circuit states, one applies a sequence of control microwave pulses to drive transitions, followed by a measurement pulse, which probes the cavity. Vanishing power in this probe tone ensures minimal back action but at the price of low signal-to-noise ratio, leading to incomplete state discrimination. This limit of ``weak measurement'' has been investigated in several works, allowing to better understand how measurement projection occurs, and to apply real-time feedback on the system \cite{Hatridge2013,Murch2013,Vijay2013,Campagne2013}. Here, we focus on the opposite limit, in which strong measurement allows unambiguous state discrimination \cite{Takmakov2021,Gusenkova2021}. The evolution of the states separation when increasing measurement power reveals the non-linearity of the system. Although this regime gives access to the transitions dynamics \cite{Blais2021,Gusenkova2021,Janvier}, we show here that it is strongly affected by the number of photons in the cavity \cite{Gusenkova2021}.

The role of the number of photons in cQED was studied theoretically by many authors, mostly considering a pure two-level system (qubit) coupled to a harmonic oscillator, at different levels of approximation \cite{Blais2021,Boissonneault,Boissonneault2009,Boissonneault2010, Bishop2010,Sete,Hanai}. The principal effects of the presence of photons in the cavity are: i) a change in the state-dependent cavity frequency shift and ii) a change in the dynamics of the qubit (including relaxation and excitation rates) \cite{Boissonneault,Boissonneault2009,Boissonneault2010,Sete,Hanai}. 
The former is a consequence of the non-linearity inherited by the cavity due to its coupling to the qubit and the latter is produced by the dressing of the intrinsic collapse operators that describe excitation and relaxation. Both effects have been reported on quantum circuits with a low anharmonicity for which the two-level truncation becomes inadequate at large power \cite{Sank2016,Cohen2023,Shillito2022,Khezri2023}.

\begin{figure}[t!]
\includegraphics[width=\columnwidth]{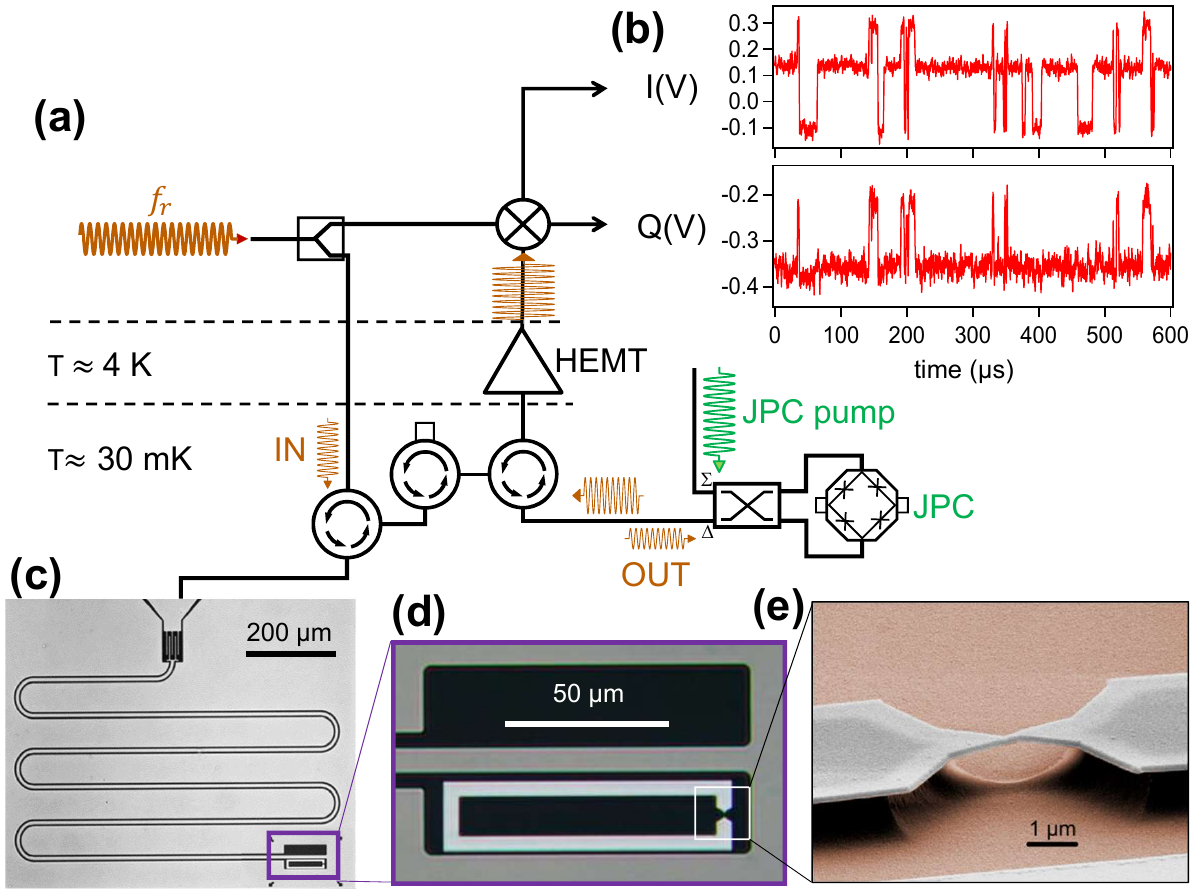}
\caption{\textbf{Continuous monitoring of the quantum state of an atomic-size contact:} (a) microwave setup: a continuous microwave tone at the frequency of the bare resonator $f_r$ is sent to the microwave cavity. The reflected signal is amplified at low temperature by a Josephson parametric converter (JPC), then mixed with a local oscillator tone to obtain the homodyne components $I$ and $Q$ of the response. One point every 10~ns is recorded with a fast acquisition card, resulting in time traces like the ones shown in (b), which display jumps between states. (c) The cavity is a coplanar quarter-wavelength microwave resonator patterned in a 150~nm-thick Nb film on a Kapton substrate. (d) An aluminum loop is fabricated near the resonator shorted end. (e) The loop contains a narrow suspended bridge that can be broken, under cryogenic vacuum, by bending the substrate. Atomic-size weak links hosting Andreev bound states are obtained by bringing the two resulting electrodes back into contact. Different atomic configurations are obtained by fine-tuning the substrate curvature.
}
\label{setup}
\end{figure}

In this work, we investigate experimentally these effects using a highly anharmonic quantum circuit: an atomic-size weak link in a superconducting loop \cite{Janvier,Metzger2021} (see Fig.~\ref{setup}). The atomic-scale weak link, in short atomic contact, hosts a few spin-degenerate Andreev levels within the superconducting gap \cite{Beenakker}. In the case of Al contacts, they can be adjusted such that one Andreev level is at an energy much smaller than the others, so that essentially only this one is probed in a cQED setup. An Andreev level can be occupied by 0, 1 or 2 quasiparticles, leading to distinct responses of the coupled cavity  \cite{Janvier, thesisJanvier, Hays2020,Metzger2021}. We performed continuous measurements of the cavity at various microwave powers, analyzed the states discrimination and extracted the transition rates. Our setup allowed us to realize, in the same cooldown, these measurements on many different contact configurations of different Andreev level energies.

The paper is organised as follows: in Section II, we describe the experimental setup. We then review and extend the theoretical predictions of cQED with a driven cavity: the effects of the cavity photon number on the qubit-dependent cavity shift are discussed in Section III and those on the dressed qubit dynamics in Section IV. In both sections we compare the corresponding measurements with theoretical predictions. In Section V, we discuss the transition rates extrapolated to the zero-photon limit in terms of Andreev physics.

\section{cQED setup with atomic contacts}
\subsection{Andreev states}
Superconducting atomic-size contacts host just a few Andreev levels \cite{Scheer1997}. Aluminum contacts can be tuned such that a single transport channel has a transmission $\tau$ very close to 1, the one or two others having moderate transmissions \cite{Janvier, thesisJanvier}. For one channel, the Andreev level at energy $E_A=\Delta_s \sqrt{1-\tau \sin^2(\varphi/2)}$ can be occupied by 0, 1 or 2 quasiparticles, which correspond to the circuit quantum states $|g\rangle$, $|o\rangle$ and $|e\rangle$ at energies $-E_A,$ $0$ and $E_A$, respectively. Here, $\Delta_s$ is the superconducting gap and $\varphi$ is the superconducting phase difference across the atomic contact. For Al, $\Delta_s/h \approx 44~$GHz. When the fermion parity is even, $|g\rangle$ and $|e\rangle$ form a two-level system (Andreev qubit \cite{Desposito2001,Zazunov2003}), with transition energy $h f_A=2 E_A.$ When the parity is odd, there is a single, spin-degenerate state $|o\rangle$ with zero energy (in finite-length weak links, this degeneracy can be lifted by spin-orbit interaction, see Ref.~\cite{Tosi}). When using a cavity of frequency $f_r \ll \Delta_s/h,$
the Andreev qubit couples significantly to it only if  $1-\tau \ll 1$ and $\varphi \approx \pi,$ so that the transition energy $2E_A=2\Delta_s \sqrt{1-\tau}$ is close to $ h f_r.$ All the data presented in this work were taken at $\varphi = \pi$.  In this regime, the weak link  realizes a simple two-level system when the fermion parity is even; when it is odd, the contact does not carry any current and is decoupled from the cavity.

\subsection{Experimental setup}
The circuit comprising the atomic contact and the microwave cavity was fabricated on a Kapton substrate \cite{Zgirski} (see Fig.~\ref{setup}(c)). Atomic contacts were obtained using the microfabricated break junction technique \cite{Ruitenbeek,Janvier}. The substrate was clamped at  one end against a microwave launcher SMA connector, while a pushing rod at the other end controlled its bending \cite{thesisJanvier}. By bending the substrate, a suspended aluminum bridge is elongated until it breaks. By gently bringing back into contact the two resulting electrodes it is possible to create and fine tune different atomic-size contacts with the same circuit. The whole bending mechanism was placed inside a series of shielding boxes (superconducting shield painted inside with carbon black, cryoperm, copper), and anchored at the mixing chamber of a wet dilution refrigerator operated at $\approx \SI{40}{\milli\K}.$ The full break junction operation is performed under cryogenic vacuum. The bridge is part of a $\SI{100}{\um} \times \SI{20}{\um}$ aluminum loop (width $\SI{5}{\um}$, thickness $\SI{0.1}{\um}$), %placed at the shorted end of%
coupled to a quarter-wavelength coplanar wave-guide resonator (the cavity) with bare resonance frequency $f_r=\omega_r/2 \pi$, see Fig.~\ref{setup}(d). The resonator itself is coupled at its other end to the measurement circuitry through an interdigitated capacitor $C_c\approx \SI{15}{\fF}.$ 
The resonator was probed in reflection by a microwave tone at frequency $f_0 \approx f_r$, amplified by a Josephson Parametric Converter (JPC) \cite{Bergeal,Roch} placed at the mixing chamber and then by a HEMT at $\SI{1.2}{\K}.$ The in-phase ($I$) and out-of-phase ($Q$) quadratures of the signal were obtained by homodyne demodulation. 
The microwave resonator (or cavity) was first characterized with the bridge fully open, giving $f_r=\SI{8.77}{\GHz},$ internal and total quality factors $Q_i=4500$ and $Q_t$=950 (total cavity decay rate $\kappa/2\pi =f_r/Q_t=9.2~$MHz). 
When a high-transmission contact is formed, measurement points acquired on periods longer than the parity switching time cluster into 3 clouds in the $(I,Q)$ plane corresponding to the pointer states associated to $|g\rangle$ (cavity displaced by the ``cavity pull'' $-\chi$), $|o\rangle$ (undisplaced cavity) and $|e\rangle$ (cavity displaced by $+\chi$). Pulsed two-tone measurements are performed to determine the transition frequency $f_A$ between $|g\rangle$ and $|e\rangle$ \cite{Janvier}.

\section{Effect of the photon number on the qubit-dependent cavity shift}
Although the state-dependent shift of the cavity frequency decreases with the measurement power $P$, we show that the signal-to-noise ratio keeps increasing with $P$. We then discuss how the dynamics of the system is modified when the number of photons in the cavity, which is at low power proportional to $P$, is increased.\\

\subsection{Steady-state number of photons in the cavity}
We consider a qubit with transition energy $\omega_q$ (in the case of Andreev qubits, $\omega_q$ is noted $\omega_A$, but the discussion here is general) coupled to a cavity at $\omega_r$ with a coupling strength $g$. Within the rotating wave approximation that assumes $|\omega_q-\omega_r| \ll \omega_q+\omega_r$, the coupled system is described by the Jaynes-Cummings Hamiltonian \cite{Blais2021} 
\begin{equation}  H_{\text{JC}}=\hbar\omega_r\left(a^{\dagger}a^{}+\frac{1}{2} \right)+\frac{\hbar \omega_q}{2} \sigma_z + \hbar g \left(a^{\dagger} \sigma_- + a \sigma_+  \right),
\end{equation}
with $a$ and $a^{\dagger}$ the standard annihilation and creation operators of the harmonic oscillator, $\sigma_+=\ket{e}\bra{g}$ and $\sigma_-=\ket{g}\bra{e}$. 
When $g\ll |\Delta|$, with $\Delta=\omega_q-\omega_r$ the detuning, the coupling term can be treated as a perturbation, and one obtains in leading order in $\lambda=g/\Delta$:
\begin{equation}
H_{\text{0,disp}}= \hbar\left(\omega_r + \chi_0\sigma_z\right) a^{\dagger}a^{} +\frac{\hbar }{2}\left(\omega_q+\chi_0\right)\sigma_z,
\end{equation}
with $\chi_0=\frac{g^2}{\Delta}$ the ``cavity pull''. The cavity frequency is  shifted by $\pm\chi_0$, the sign depending on the qubit state. In this limit, known as the dispersive regime, driving the cavity with a tone at $\omega_0=\omega_r+\delta$ and a power $P$, leads in a steady state, as long as the qubit remains in the same quantum state, to an average photon number (see Appendix \ref{app:ddrop}):
\begin{equation}
    \overline{n}_{\text{disp}} =\frac{A_0^2}{4 \left(\delta \pm\chi_0\right)^2+\kappa^2},
    \label{ndisp}
\end{equation}
where $A_0=\sqrt{4 \kappa_{c} P/\hbar \omega_r}$ is the amplitude  of the cavity drive, with $\kappa_{c}$ the coupling decay rate \cite{Aspelmeyer2014}.

From the qubit point of view, in addition to the Lamb shift $\chi_0$, it also experiences a Stark shift $2\chi_0 a^{\dagger}a^{} $. The above approximation becomes invalid when the number of photons increases, otherwise this Stark shift could exceed the qubit energy itself. In fact, it breaks down earlier, as can be seen when developing the coupling term to the next order in $\lambda$ \cite{Boissonneault}:
\begin{equation}
\begin{split}
H_{\text{0,disp2}}= &\hbar\left[\omega_r - \zeta+ \left(\chi_0 (1-\lambda^2) -\zeta a^{\dagger}a^{}\right) \sigma_z\right] a^{\dagger}a^{}\\
&+\frac{\hbar }{2}\left(\omega_q+\chi_0\right)\sigma_z,
\end{split}
\end{equation}
where $\zeta=\frac{g^4}{\Delta^3}=\chi_0 \lambda^2$ is a Kerr-type non-linearity \cite{Boissonneault2010}.
By evaluating the energy difference between states $\ket{g,n+1}$ and $\ket{g,n}$, one obtains the dependence of the cavity pull on $n$, the number of photons in the cavity:
\begin{equation}
  \chi(n)\approx\chi_0 (1-2\lambda^2 n)=\chi_0\left(1-\frac{n}{2 n_{\rm crit}}\right),
\end{equation}
with $n_{\rm crit}=\left(\frac{\Delta}{2g}\right)^2$ the critical photon number \cite{Boissonneault2010}. This expression was extended to arbitrary ratio $\frac{n}{2 n_{\rm crit}}$ by exact diagonalization \cite{Bishop2010}, yielding 
\begin{equation}
\label{chiofn}
    \chi(n) = \frac{\chi_0}{\sqrt{1+n/n_{\rm crit}}}.
\end{equation}
Since the cavity pull depends on $n$, the average number of photons in the driven cavity $\overline{n}$ is the solution of 
\begin{equation}
    \overline{n} =\frac{A_0^2}{4 \left(\delta\pm\chi(\overline{n})\right)^2+\kappa^2}.
    \label{ndisperse}
\end{equation} 
Focusing on the experimental situation $\delta=0,$ we rewrite this equation as:
\begin{equation}
    \overline{n} =
     \frac{n_0}{1+(2 \chi_0 / \kappa)^2/(1+\overline{n}/n_{\rm crit})},
\label{n_vs_ain}
\end{equation}
where $n_0=(A_0/\kappa)^2\propto P$ is the average number of photons in the unshifted cavity, \textit{i.e.} when the qubit is in the odd state. The analytical solution of this equation gives $\overline{n}$ as a function of $n_0$. At small power, $\overline{n} \propto n_0,$ but grows faster when  approaching $n_{\rm crit}$ because the difference $\pm \chi(\overline{n})/2\pi$ between the frequency of the readout tone and that of the dressed cavity diminishes with $\overline{n}$.

\subsection{State discrimination}
We consider a cavity measured in reflection. When a tone $A_{\rm in} {\rm e}^{i \omega_0 t}$ is sent to the cavity, the reflected signal is $A_{\rm out} {\rm e}^{i \omega_0 t}=S_{11} A_{\rm in} {\rm e}^{i \omega_0 t}.$ The reflection coefficient $S_{11}$ reads \cite{thesisJanvier}
\begin{equation}
S_{11}(x)=1-\frac{Q_t}{Q_e}\left[1+\exp\left(-2 i \arctan\left(2 Q_t x\right)\right)\right]
\label{S11}
\end{equation}
with $x=\frac{\omega_0}{\omega_r^*}-1$, and $Q_e$, $Q_i$ and $Q_t=(Q_e^{-1}+Q_i^{-1})^{-1}$, the external, internal and total quality factors, respectively, and $\omega_r^*$ is the state-dependent resonance frequency.

\begin{figure}[t!]
\includegraphics[width=\columnwidth]{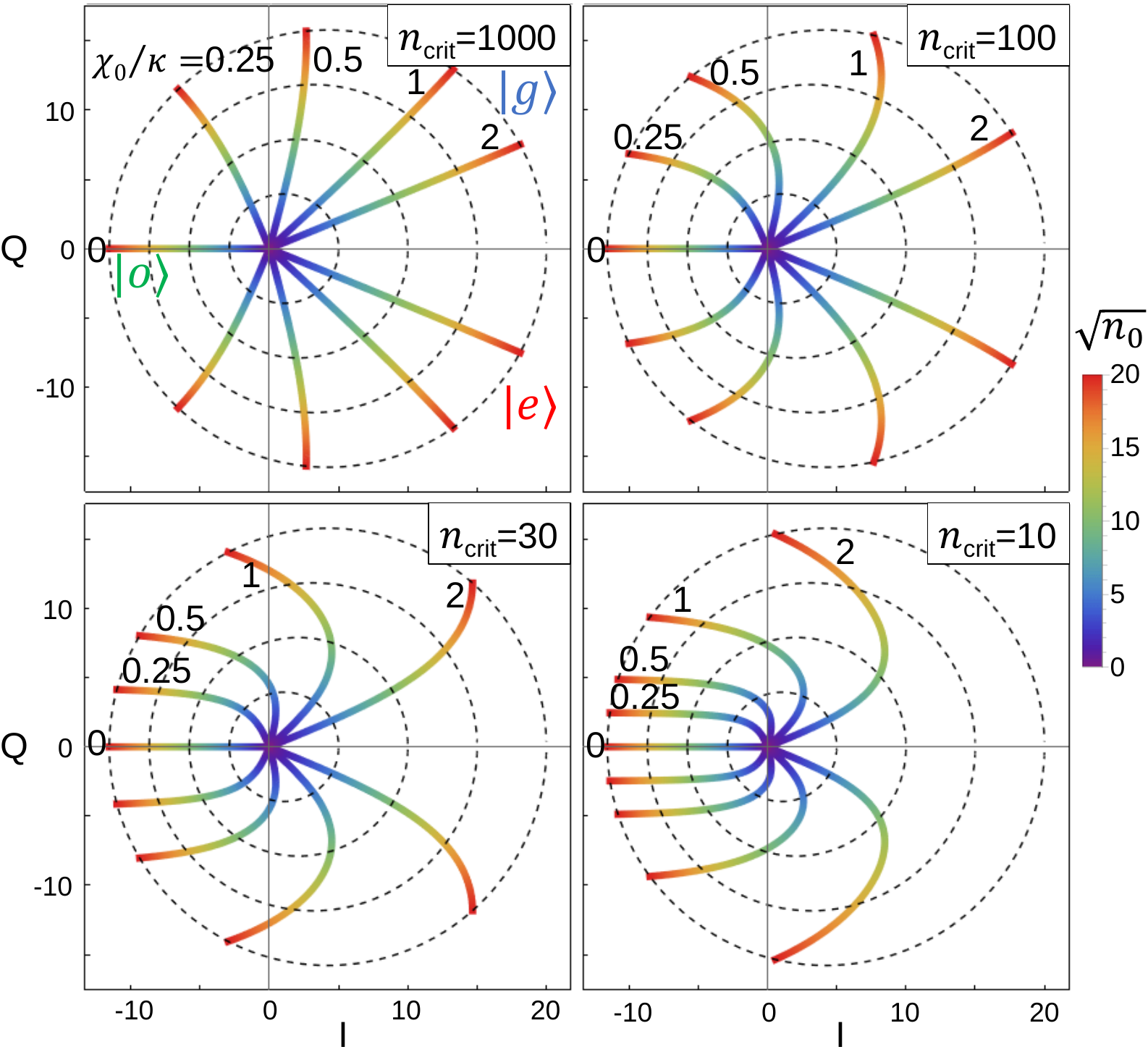}
\caption{\textbf{Pointer states positions vs probe tone amplitude:} Predicted position of the center of the pointer states corresponding to $|g\rangle$ and $|e\rangle$, on both sides of the horizontal axis, and $|o\rangle$ ($\chi_0/\kappa=0$), as a function of the normalized probe tone amplitude $\sqrt{n_0}$ (color scale), and for $n_{\rm crit}=1000, 100, 30, 10$. The color curves correspond to $\chi(0)/\kappa=0,$ $0.25,$ $0.5,$ $1,$ and $2.$ Circles correspond to $S_{11}(f)$ for $\sqrt{n_0}=5, 10, 15, 20$, and using $Q_t=950,$ $Q_i=4500.$ The panel at $n_{\rm crit}=1000$ is the closest to the linear regime: the position of the clouds evolves almost linearly with $\sqrt{n_0}$.}
\label{blobs_vs_n}
\end{figure}

The quadratures of the reflected signal are $I=\Re (A_{\rm out})$ and $Q=\Im (A_{\rm out})$. When probing the cavity at its bare frequency, $\omega_0=\omega_r$, the positions of the pointer states of the cavity in the $(I,Q)$ plane  \cite{Blais2021} are given by $\sqrt{n_0} S_{11}(\pm\chi(\overline{n}(n_0))/\omega_r)$. For the Andreev qubit, we have, in addition to the states $\ket{g}$ and $\ket{e}$, the state  $\ket{o}$ that does not shift the cavity: its pointer state is at $\sqrt{n_0} S_{11}(0)$. In Fig.~\ref{blobs_vs_n} we show the expected positions for the pointer states of $|g\rangle$, $|e\rangle$ and $|o\rangle$ as a function of $n_0$ and for various values of $\chi_0/\kappa$ and $n_{\rm crit}$. In the linear regime corresponding to $n_0 \ll n_{\rm crit}$, the states separation increases linearly with $\sqrt{n_0}$ (see first panel). When $n_0$ approaches $n_{\rm crit}$ (see Eq.~(\ref{n_vs_ain})), $\overline{n}/n_{\rm crit}$ is of order 1, and the decay of $\chi$ with $\overline{n}$ tends to group the pointer states near the leftmost point of the circle $S_{11}(x).$ Finally, when $\overline{n} \gg n_{\rm crit},$ the separation between the pointer states of $\ket{g}$ and $\ket{e}$ saturates at $4\frac{Q_t}{Q_e}\frac{g}{\Delta}.$

\begin{figure}[t!]
\includegraphics[width=\columnwidth]{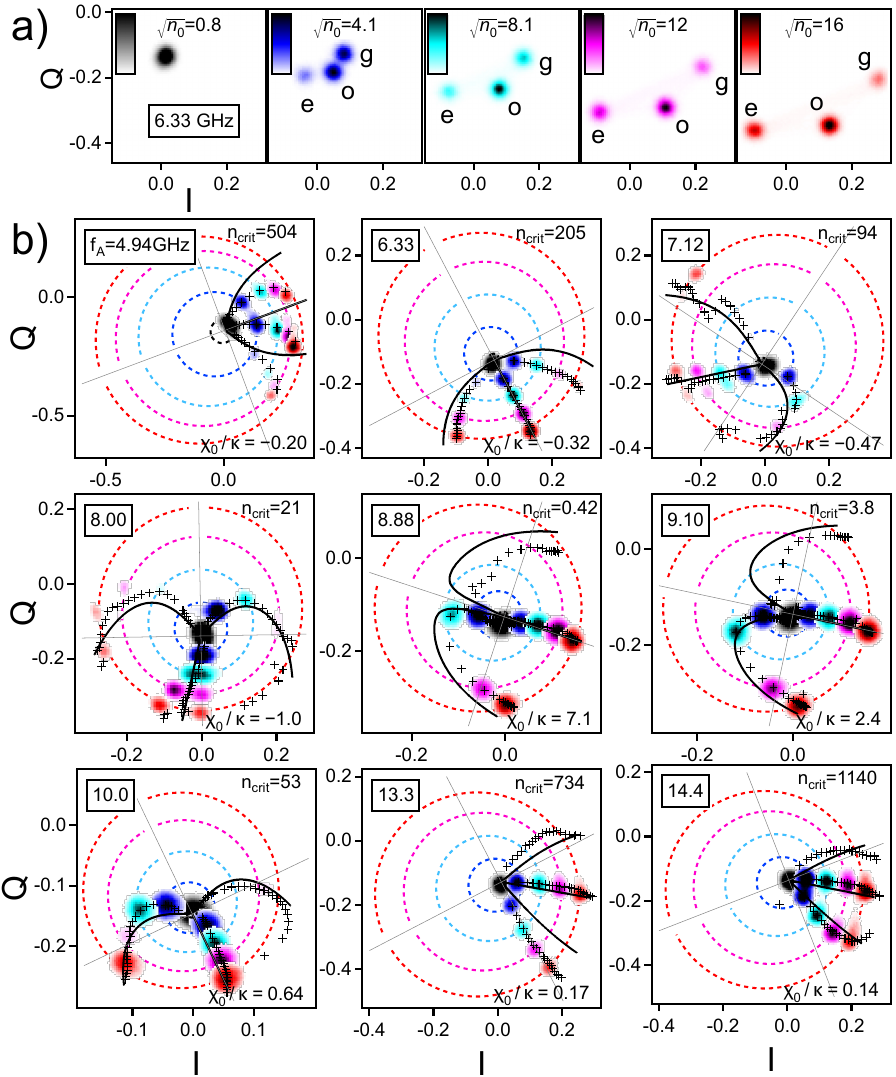}
\caption{\textbf{Histograms of the values of $(I,Q)$ recorded during continuous measurements at various probe amplitudes.} a)~Contact with $f_A=6.33~$GHz, at increasing probe amplitudes $\sqrt{n_0}$. In each panel, the color is associated to the value of $\sqrt{n_0}$, and the brightness in the color scale encodes the number of counts per pixel, with a linear variation from 0 to 12000. b) Panels correspond to 9 different contacts, with $f_A$ given in GHz in the text box. Colors correspond to increasing probe amplitudes as in (a): $\sqrt{n_0}\approx$ $1$ (black), $4$ (blue), $8$ (cyan), $12$ (magenta) and $16$ (red) (precise values in Appendix, Table I). Color brightness scale limits are set between max/6 and max to reduce histogram overlaps. Dashed circles correspond to predictions for $\sqrt{n_0} S_{11}(f)$, globally scaled for each contact in order to account for the measurement gain, shifted and rotated to align with the clouds. Crosses indicate the positions of the center of the 3 clouds inferred from the analysis of the traces, at all values of $\sqrt{n_0}.$ Solid lines are predicted positions $\sqrt{n_0} S_{11}(\pm\chi(\overline{n}(n_0))+\delta \omega),$ and $\sqrt{n_0}S_{11}(\delta \omega),$ scaled, rotated and shifted as the circles. A constant frequency shift $\delta \omega$ was adjusted in each series, see Table~I.}
\label{histos}
\end{figure}

\subsection{Clouds in the $(I,Q)$ plane}
We acquired 1-s-long continuous measurement time traces of $I$ and $Q$ (one point every 10~ns), as illustrated in Fig.~\ref{setup}(b), at different amplitudes of the probe tone. Figure~\ref{histos}(a) shows histograms of the values of $(I,Q)$ from 5 such traces after a box-averaging over 1~$\mu$s, for an atomic contact with $f_A=6.33$~GHz, at increasing normalised probe amplitudes $\sqrt{n_0}$. Except at the lowest amplitude where they are superimposed, the three clouds corresponding to $\ket{g},$ $\ket{o}$ and $\ket{e},$ get apart when $n_0$ increases. 
Figure~\ref{histos}(b) shows similar measurements for nine different contacts, with $f_A$ ranging from 4.94 to 14.4~GHz. The non-linearity associated with the reduction of $\chi$ with $\overline{n}$ is revealed by the bending of the trajectories described by the clouds.
In Fig.~\ref{histos2} in Appendix \ref{app:clouds}, we show  histograms at different driving powers obtained from  20 to 40 traces concatenated, to give an alternative view of the evolution of the clouds.
 \begin{figure*}[t!]
\includegraphics[width=1.8\columnwidth]{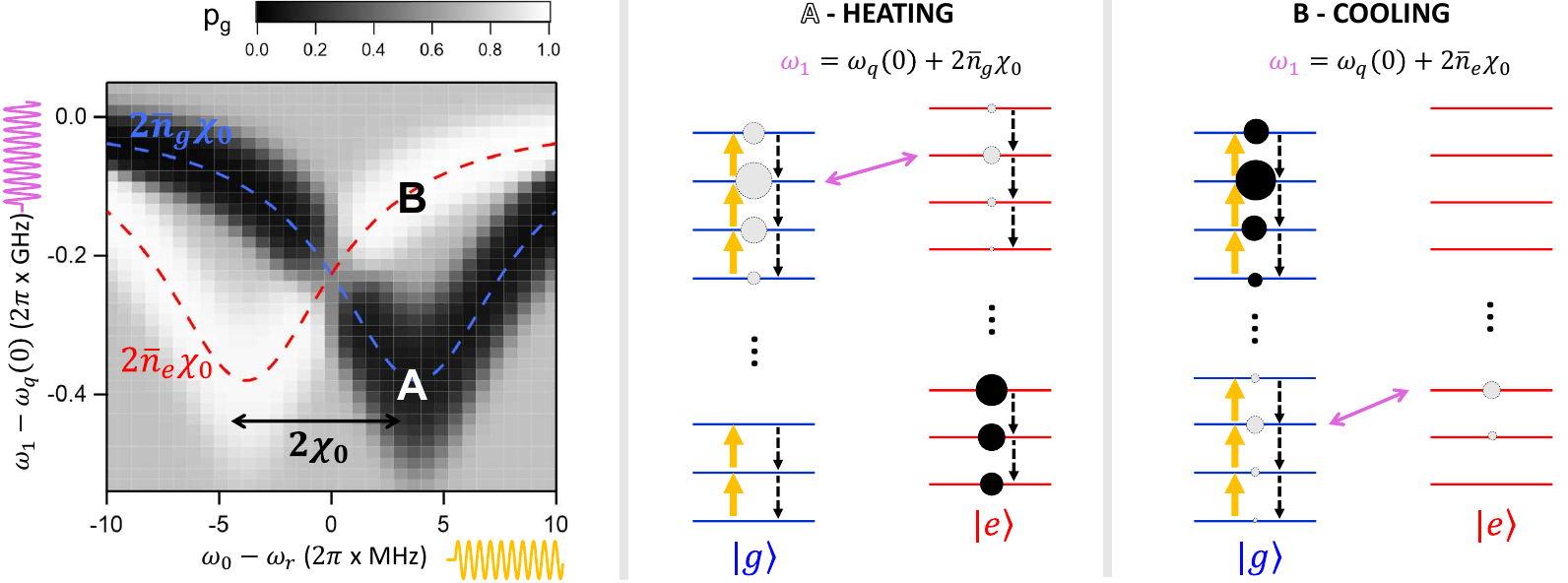}
\caption{\textbf{Schematic explanation of a DDROP (double drive reset of population) measurement}. The protocol consists in performing the qubit 2-tone spectroscopy in presence of a tone at frequency $f_0=\omega_0/2\pi$ close to that of the cavity. Left: steady state population of the ground state of the qubit calculated in the dispersive approximation, as a function of $\omega_0$, and of drive frequency $\omega_1.$ According to Eq.~(\ref{ndisp}), the average number of photons in the cavity $\overline{n}_g$ is described by a lorentzian with a maximum at $\omega_0-\omega_r=-\chi_0~(>0)$, when the qubit is in its ground state, and a mirrored lorentzian centered at $\omega_0-\omega_r=\chi_0~(<0)$ for $\overline{n}_e$ corresponding to the excited state. The blue (red) dashed line corresponds to $\omega_1-\omega_q(0)=2\overline{n}_g \chi_0$ ($=2\overline{n}_e \chi_0$) and describes the Stark shift of the qubit transition.  Panels \textbf{A} and \textbf{B} correspond to the qubit driven at $\omega_1=\omega_q (0)+2\overline{n}_g\chi_0~$ and $\omega_1=\omega_q (0)+2\overline{n}_e\chi_0~$ respectively.  In both cases, a cavity drive at $\omega_0=\omega_r-\chi_0~$ creates a coherent state in which only some upper levels of the $|g\rangle$ ladder are significantly populated (populations represented with disks). In A, the qubit is driven  at its shifted frequency, $\omega_1-\omega_q (0)=2\overline{n}_g \chi_0$ giving rise to Rabi oscillations between the two ladders around the $\overline{n}_g$-th level (populations of involved levels shown with grey disks). As in the qubit excited state $|e\rangle$ the cavity drive is not resonant, cavity decay transfers the population to the lowest states of the ladder (black disks). This leads to a steady state with an accumulation of population in the excited state (heating). In \textbf{B}, the qubit drive is resonant with $\omega_1-\omega_q (0)=2\overline{n}_e \chi_0$, giving rise to Rabi oscillations between the two ladders around the $\overline{n}_e$-th level (populations of involved levels shown with grey disks). Here, the cavity drive, which is resonant when the qubit is in $|g\rangle$,  transfers the population to higher energy levels of the $|g\rangle$ ladder. As a result, the population of $|g\rangle$ becomes larger than at thermal equilibrium (cooling). \textit{Simulation parameters:} $f_q=$7.15~GHz, $g/2\pi=0.078$~GHz, $\overline{n}=$50, $f_r=8.77~$GHz and $\kappa/2\pi=9.2~$MHz.}
\label{ddrop}
\end{figure*}

To compare the data of Fig.~\ref{histos}(b) with Fig.~\ref{blobs_vs_n}, we overlaid circles corresponding to the functions $\sqrt{n_0} S_{11}(f)$ for the 5 values of $\sqrt{n_0}$. The relative values of $n_0$ correspond to the power settings of the microwave source; the absolute values result from the calibration described in section \ref{DDROP}. Comparison requires a shift of the origin, to account for the offsets of the amplifiers, a rotation to account for the measurement phase, and  a global scaling corresponding to the gain of the JPC that varies from one set of data to the other since it was optimized for each contact. For each dataset, these parameters were manually adjusted so that the clouds at the different amplitudes fall on the corresponding circles. For some of the contacts, more than 3 clouds can be distinguished when the measurement amplitude is large, in particular at 7.12 and 8.0~GHz, indicating the presence of a second Andreev state at a higher energy, with correspondingly a smaller $\chi_0$, and measured either in its ground or in its odd state.
Despite this complication, all traces were analysed assuming the presence of 3 clouds \cite{SMART}, and the positions of the centers of the clouds at all values of $n_0$ are indicated with crosses in Fig.~\ref{histos}.

In order to compare the changes in the clouds position with theory, we calibrated the power of the measurement tone and the photon number, as explained now.

\begin{figure}
\includegraphics[width=0.95\columnwidth]{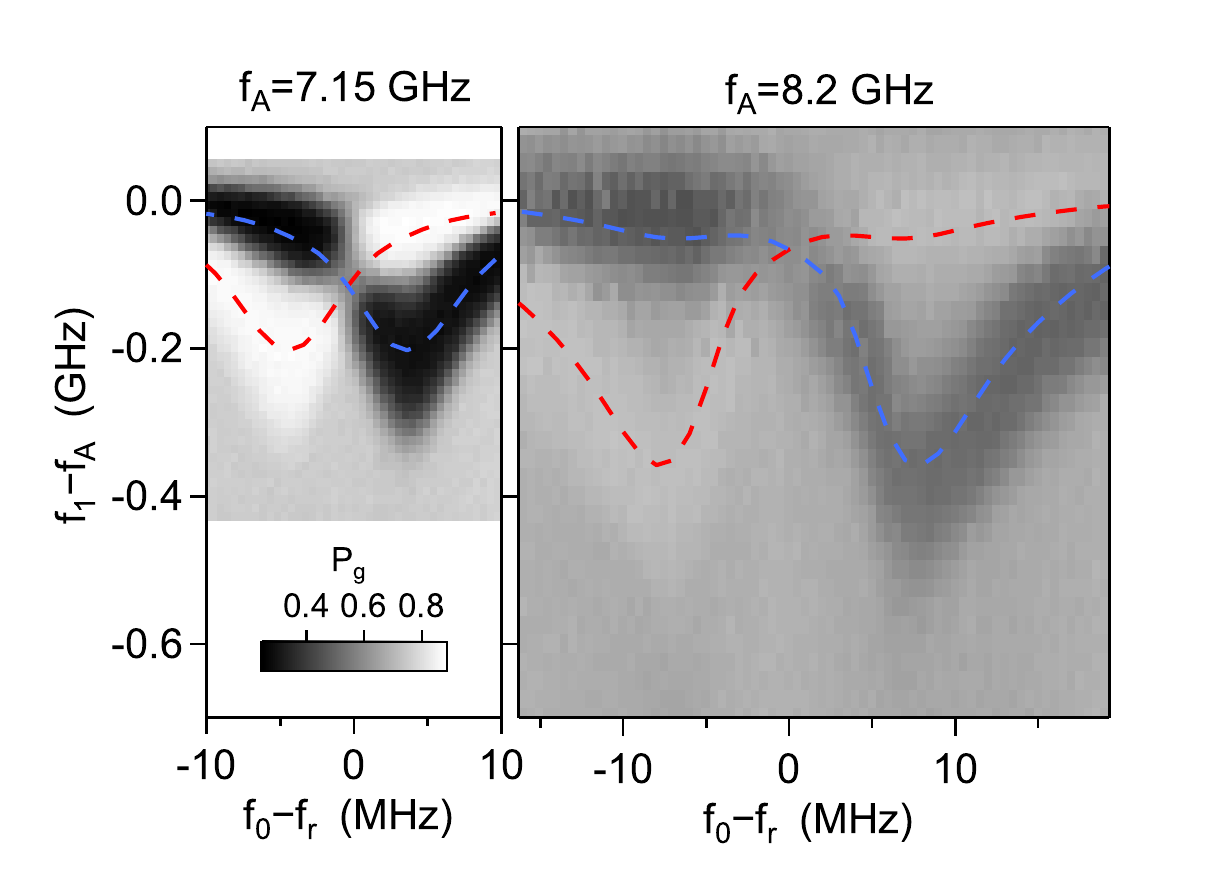}
\caption{DDROP measurements for qubits at 7.15 and 8.2~GHz using a cavity drive amplitude corresponding to $\overline{n}=25$ and $\overline{n}=22$ photons respectively (see discussion on calibration in the text). The qubit and cavity drive pulses are applied simultaneously during 10~$\mu$s followed by a 1~$\mu$s measurement pulse at $f_r=8.77~$GHz. The grey scale corresponds to the population $p_g$ of the ground state. The smaller contrast at $f_A=8.2~$GHz is attributed to a shorter lifetime than at 7.15~GHz. The dashed lines correspond to Eq.~(\ref{fqtheo}). While for 7.15~GHz the experiment resembles Fig.~\ref{ddrop}, a simplified dispersive theory would predict an erroneous qubit shift. The effect is more pronounced when the qubit frequency approaches the cavity frequency, as one observes non-Lorentzian resonances arising from the non-linearity.}
\label{ddrop2}
\end{figure}

\subsection{Calibration with DDROP protocol}\label{DDROP}

The method of choice to calibrate the photon number is to measure the induced dephasing \cite{Gambetta}. In the case of atomic contacts, the typical coherence time $T_2$ is of the order of 20~ns so this procedure could not be applied. Instead, we have performed 3-tone measurements following the DDROP protocol (``double drive reset of population'' \cite{Geerlings2013}). It consists in performing the spectroscopy of the qubit (drive tone at $f_1=\omega_1/2\pi$) in presence of a tone at $f_0=\omega_0/2\pi=(\omega_r+\delta)/2\pi$ close to the bare cavity frequency (see Fig.~\ref{ddrop}). The saturating tones at $\omega_0$ and $\omega_1$ are applied simultaneously, and followed by a measurement pulse. The principle of the calibration, designed for the dispersive limit, is that the qubit transition frequency follows $\omega_q(n)=\omega_q(0)+2\overline{n}\chi_0$. 
Let us assume that $\chi_0<0$. Due to the cavity pull $\pm \chi_0$, the cavity photon number takes different values $\overline{n}_g$ and $\overline{n}_e$ if the qubit is in $\ket{g}$ or $\ket{e}$ respectively. When $f_0$ matches the resonator frequency for the qubit in its ground state ($\omega_0=\omega_r-\chi_0$), $\overline{n}_g$ reaches a maximum, and the transition frequency presents a minimum at $\omega_1=\omega_q(0)+2\overline{n}_g \chi_0$. In this situation, the Rabi drive associated to the drive tone at $\omega_1$ combined with photon decay when the qubit has some weight in $\ket{e}$ leads to an excess population in the excited state (see central panel in Fig.~\ref{ddrop}). 
The steady state obtained when the qubit is driven at $\omega_1=\omega_q(0)+2\overline{n}_e \chi_0$ is illustrated in the rightmost panel of Fig.~\ref{ddrop}: when the qubit is in $\ket{e},$ the Rabi drive  combined with the tone at $\omega_0$, resonant with the cavity when the qubit is in $\ket{g}$, transfers population to $\ket{g}$. A symmetric behavior occurs when $\omega_0=\omega_r+\chi_0,$ leading, in a plot of the population $p_g$ of $\ket{g}$ as a function of $f_0$ and $f_1$, to two lorentzian dips shifted by $2\chi_0$ and with an amplitude $2 \overline{n} \chi_0$. The dip separation provides a calibration of $\chi_0$. The ratio of amplitude and separation of the dips calibrates $\overline{n}$ for a resonant cavity drive, at the amplitude chosen for $f_0$.
A quantitative description is obtained using  the Lorentzian variations of $n$ with $\delta$ (Eq.~(\ref{ndisperse}), in which the variations of $\chi$ with $n$ are neglected):
\begin{equation}
    \omega_q(n(\delta))-\omega_q(0)=
    \frac{2 n_0 \chi_0}{4 \left(\frac{\delta+\chi_0}{\kappa}\right)^2+1}.
\end{equation}

One obtains the curves shown with blue and red dashed line in Fig.~\ref{ddrop}. The grey-scale map is a numerical simulation of the steady-state population of the qubit obtained with the Qutip package \cite{QuTip} for a Jaynes-Cummings Hamiltonian in the dispersive limit including two drives (see App. \ref{app:ddrop}).   

Experimental results at two values of $f_A$ are shown in Fig.~\ref{ddrop2}. The population $p_g$ and $p_e$ of the ground and excited state (normalised to $p_g+p_e$) were extracted from the measured average quadratures $\bar{I},\bar{Q}$, using the positions $(I_i,Q_i)$ of the clouds corresponding to states $i=g,o,e$, and the equations $\bar{I}=\sum_i p_i I_i,$ $\bar{Q}=\sum_i p_i Q_i$, and $\sum_i p_i =1$. 

\begin{figure*}[t!]
\includegraphics[width=2\columnwidth]{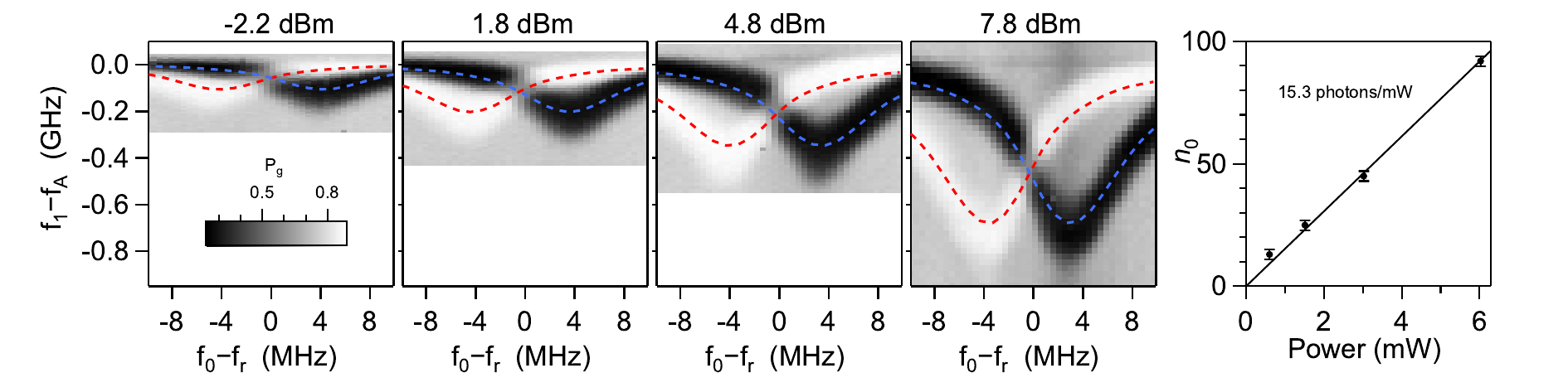}
\caption{DDROP measurement of a contact with $f_A\approx7.15~$GHz with increasing cavity drive power. Dashed lines are fits, yielding the dependence of cavity mean occupation $\overline{n}_g$ (blue) and $\overline{n}_e$ (red) according to Eq.~(\ref{n_vs_ain}), hence the photon number at resonance $n_0$ shown in the rightmost panel (during data acquisition, $f_A$ slightly drifted: 7.15~GHz for the first two panels, then 7.12~GHz, and 7.2~GHz for the last one).}
\label{ddropcalib}
\end{figure*}

Whereas at $f_A=7.12~$GHz the data resemble qualitatively the simulation of Fig.~\ref{ddrop2}, these at $f_A=8.2~$GHz show very asymmetric dips in the position of the resonances. This is due to the non-linearity of the resonator when coupled to the qubit:
what is measured is in fact $\chi(\overline{n})$. 
For the cavity, the key idea to calculate the shift, is that in the Fock state $n=0$, the transition frequency is $\omega_q+\chi_0$; in $n=1$, it is $\omega_q+2\chi_0+\chi(1)$; in $n=2$, it is $\omega_q+2\chi_0+2\chi(1)+\chi(2)$ and so on. That is
\begin{equation}
    \omega_q(\hat{\rho})=\omega_q+ \sum_{n=0} \left[\chi(n) + 2\sum_{i<n}\chi(i)\right]P_{\hat{\rho}}(n),
    \label{fqtheo}
\end{equation}
with the steady-state matrix density $\hat{\rho}$ giving the probability distribution $P_{\hat{\rho}}(n)$. 
We simulated the full Jaynes-Cummings Hamiltonian with the driven cavity for different values of $\delta$ and calculated the shift with Eq.~(\ref{fqtheo}) (see App. \ref{app:ddrop}).
The calibration of $g$ results from a comparison of these simulations with data taken at various drive powers on a contact with $f_A \approx 7.15$~GHz, see Fig.~\ref{ddropcalib}.  The blue and red dashed lines in Fig.~\ref{ddropcalib} show the results obtained with the best value $g/2\pi=85$~MHz, and with the photon numbers indicated in the rightmost panel. Dashed lines in Fig.~\ref{ddrop2} use the same $g$ and the photon numbers given in the caption that result from the calibration.

\begin{figure}
\includegraphics[width=0.8\columnwidth]{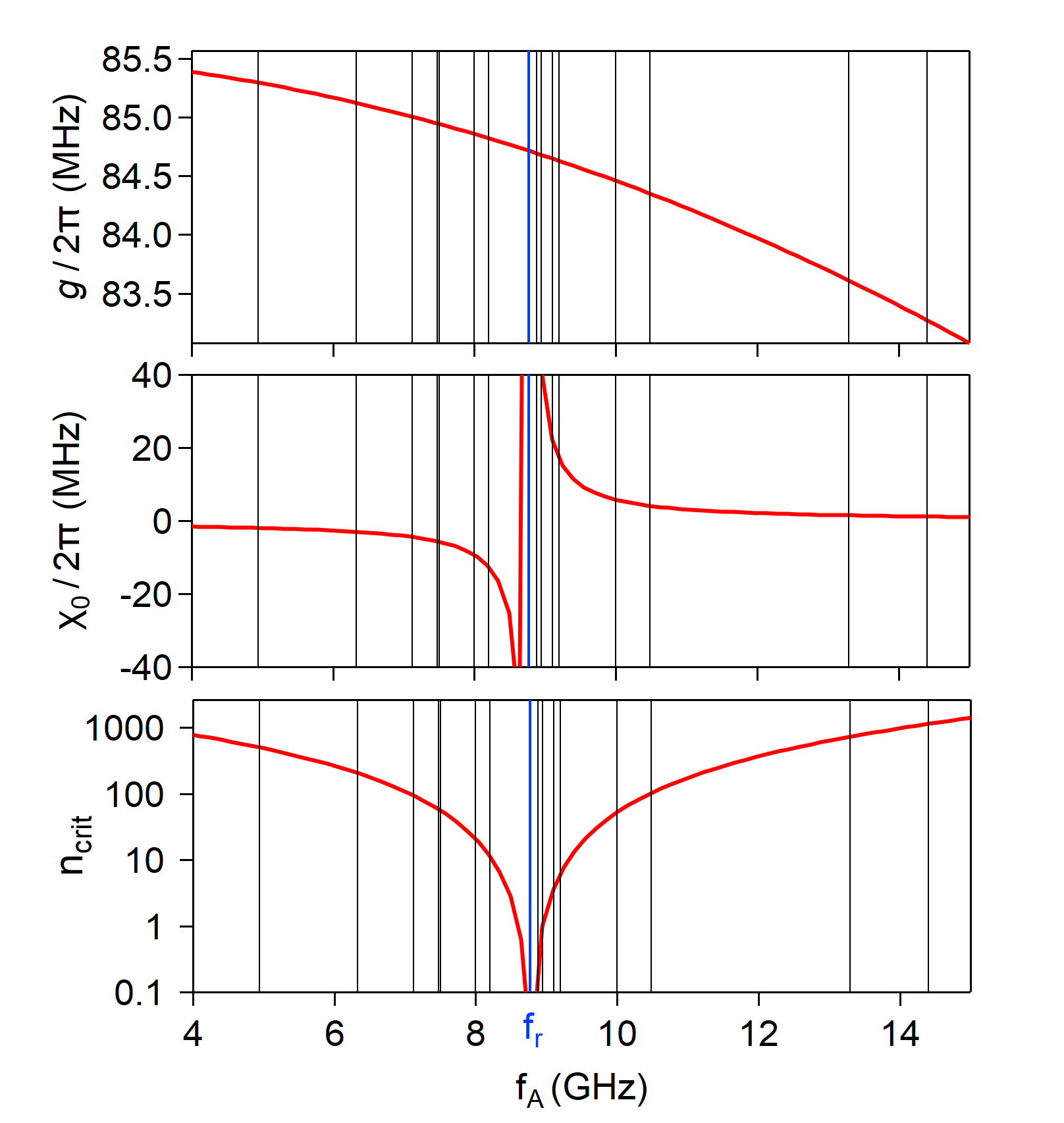}
\caption{Dependence on Andreev frequency $f_A$ of the coupling $g$, the cavity pull $\chi_0$ and the critical photon number $n_{\rm crit}.$ The black vertical lines correspond to the Andreev frequencies of the measured atomic contacts. The blue line corresponds to the resonator bare frequency $f_r$.}
\label{gchincrit}
\end{figure}

\subsection{Comparison of clouds positions with theory}
To compare clouds position with theory, we calculated for each dataset the values of $n_{\rm crit}$ and $\chi_0$, and computed the functions $\sqrt{n_0}S_{11}(\pm\chi(\overline{n}(n_0))+\delta \omega),$ and $\sqrt{n_0}S_{11}(\delta \omega),$ with the scaling, rotation and shifts obtained by aligning the clouds with the circles corresponding to $\sqrt{n_0}S_{11}(f)$ (see section C). 
We used the value of $g/2\pi=85\ $MHz obtained at 7.15~GHz, and took into account the small dependence of $g$ with $E_A$: $g \propto \left(\Delta_{\rm sc}-E_A^2/\Delta_{\rm sc}\right)$ \cite{Janvier}. The corresponding dependencies of $g,$ $\chi_0$ and $n_{\rm crit}$ on $f_A$ are displayed in Fig.~\ref{gchincrit}.
In each dataset of Fig.~\ref{histos}, the offset $\delta \omega$ was adjusted to improve comparison with theory, see Table I in Appendix \ref{app:clouds}. Such an offset can occur if additional channels at higher energies cause an additional cavity shift \cite{Metzger2021}, which can be assumed independent of $n$.

The results are shown with solid lines in Fig.~\ref{histos}. The overall change of the clouds positions with photon number is well captured. Nevertheless, the agreement is not quantitative. It is partly due to the underestimation of $\chi_0$ in the rotating wave approximation (RWA), which, for the largest detunings, can be as large as 25\% compared the result obtained beyond RWA \cite{Park2020}:
\begin{equation}
\chi_0=g^2\left(\frac{1}{\omega_q-\omega_r}+\frac{1}{\omega_q+\omega_r}\right).
\end{equation}
Further work is however needed to describe the variations of $\chi$ with the photon number beyond RWA. Other effects like the presence of additional Andreev states with higher energy can also lead to discrepancies.

In addition to the position of the clouds, an observation can be made on their relative populations. At low amplitude, the clouds corresponding to $\ket{g}$ and $\ket{o}$ are more populated than that of $\ket{e}$. But the relative population of $\ket{g}$ and $\ket{e}$ vary with $n_0,$ leading in some cases to a population inversion, as best illustrated in Fig.~\ref{histos}(a): the top right cloud, associated to $\ket{g}$, is more populated than the bottom left one ($\ket{e}$) at $\sqrt{n_0}=4.1$, but this inverts for $\sqrt{n_0}=12$ and $16$. As we will see in Section~IV, this can be related to the variations with $\overline{n}$ of the transition rates between states. 

\section{Effect of photon number on dressed qubit dynamics}
\subsection{Rates renormalization}\label{Rates renormalization}
Having explained how the coupling between the qubit and the resonator leads to non-linearities in the power dependence of the clouds positions and photon number, we discuss now how the dynamics of the qubit is affected by the presence of photons in the cavity. The main effects have been described in a series of papers by M.~Boissonneault, J. Gambetta and A. Blais \cite{Boissonneault, Boissonneault2009,Boissonneault2010}, in the limit $n \ll n_{\rm crit}.$ Since our experiments go beyond this limit, we rederived and extended their results.

We consider that the uncoupled qubit ($g=0$) is in contact with a bath that causes relaxation, excitation and dephasing at rates $\Gamma^{0}_\downarrow$, $\Gamma^{0}_\uparrow$ and $\Gamma^{0}_\phi$, respectively. In the Lindblad equations that describe the time-evolution of the qubit density matrix (see Appendix \ref{app:fit}), these processes are taken into account by the corresponding collapse operators $\sigma_-$, $\sigma_+$ and $\sigma_z$. While relaxation and excitation rates are related to noise spectral density $S_{\perp}(\omega)$ of bath fluctuators that couple to $\sigma_x$ and $\sigma_y$, pure dephasing is associated to terms proportional to $\sigma_z$ with a coefficient $S_{\parallel}(\omega)$. The thermalization with the bath determines the thermal population of the qubit $p_{e,th}=\Gamma^{0}_\uparrow/\left(\Gamma^{0}_\uparrow+\Gamma^{0}_\downarrow\right)$. 

In turn, the cavity is thermalized with its own bath by the emission of photons at a rate $\kappa$ and the absorption at a rate $\kappa {\rm e}^{-\hbar \omega_r/k_BT}$. In the equations for the time-evolution of the cavity, this is described by the action of the annihilation and creation operators $a$ and $a^{\dagger}$. The rates are proportional to the noise spectral density of the electromagnetic environment  $S_{\kappa}(\omega)$ at frequencies $\pm \omega_r$. 

\begin{figure*}[t!]
\includegraphics[width=2\columnwidth]{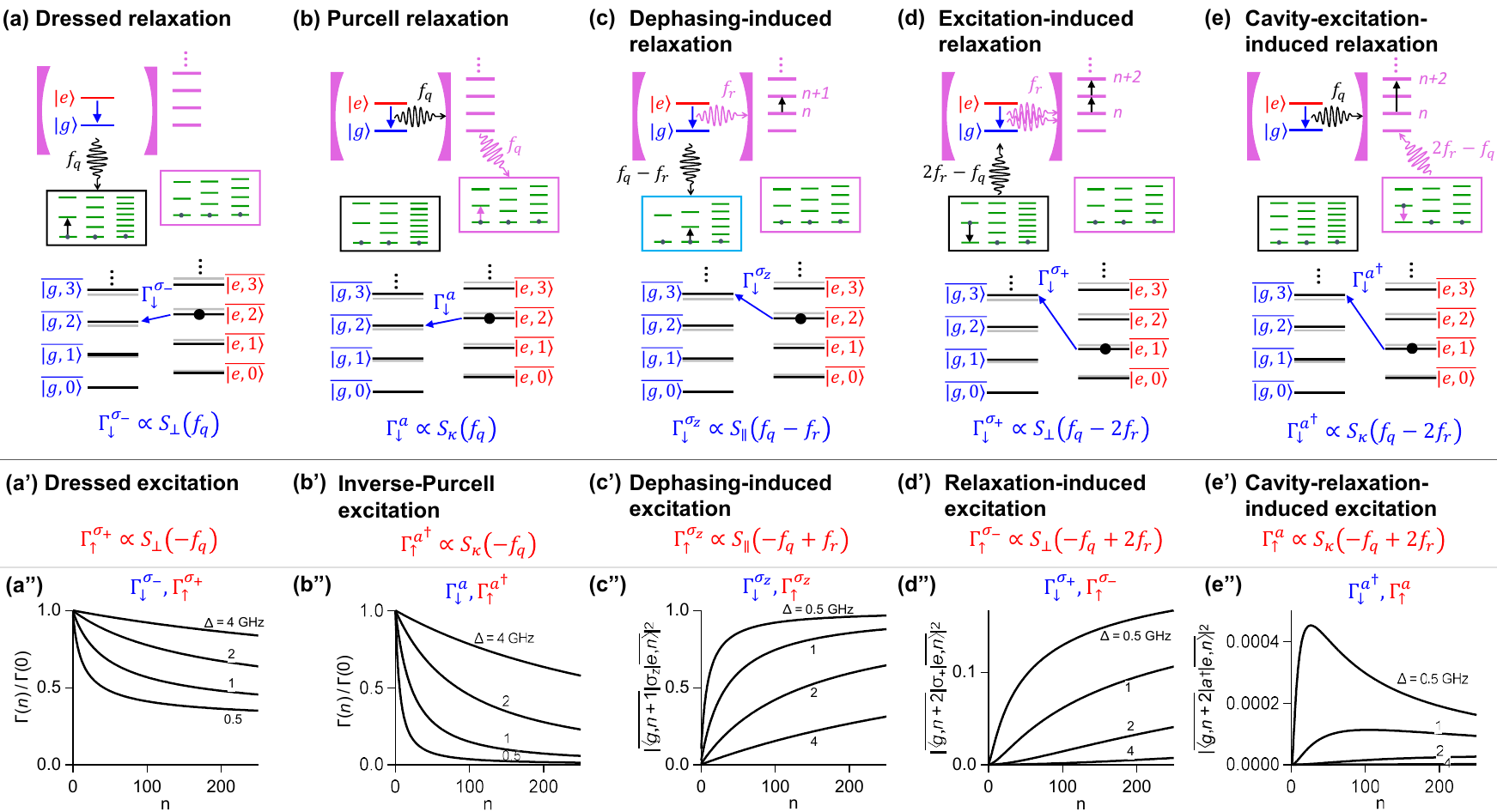}
\caption{\textbf{Contributions to the dressed dynamics:} Each panel shows the qubit levels $\ket{g}$ and $\ket{e}$ (blue and red), the cavity and cavity levels (magenta). Black and cyan rectangles represent the baths that couple to the qubit through $\sigma_{x,y}$ and $\sigma_{z}$, respectively. Magenta rectangle correspond to a bath coupled to the cavity. Bath energy levels are shown in green as a collection of harmonic oscillators. The two shifted ladders below schematize the energies of the combined states $\ket{g,n}\equiv \ket{g}\otimes\ket{n}$ and $\ket{e,n}\equiv \ket{e}\otimes\ket{n}$ (grey) and dressed states $\overline{\ket{g,n}}$ and $\overline{\ket{e,n}}$ (black). 
(a) Dressed, (b) Purcell, (c) dephasing-induced, (d) excitation-induced and (e) cavity-excitation-induced relaxation. The direction of the photon wavy arrows assume $f_r<f_q<2f_r.$
(a') to (e') are the corresponding excitation processes, with all the arrows pointing in the opposite direction.
(a'') to (e'') Dependence of the rates as a function of the cavity occupation $n$, for different values of the detuning $\Delta$ (0.5, 1, 2, 4 GHz), and using the experimental parameters ${g}/{2\pi}=85~\rm{MHz}$ and $f_r=8.77$~GHz.}
\label{rates_renorm}
\end{figure*}

When qubit and cavity are coupled, they become entangled and all collapse operators contribute to the different rates, as explained below.
An insight into this can be obtained from the expression of the dressed states of the coupled cavity-qubit system in the RWA \cite{Blais2021}:     
\begin{align}
\overline{\ket{g,n}}&=c_n\ket{g}\otimes\ket{n}+s_n\ket{e}\otimes\ket{n-1},\nonumber\\ \overline{\ket{e,n}}&=c_{n+1}\ket{e}\otimes\ket{n}-s_{n+1}\ket{g}\otimes\ket{n+1},
\label{dressedstates}
\end{align}
where $c_n=\cos \theta_n,$ $s_n=\sin \theta_n, $ and $\theta_n=\frac{1}{2}\arctan\left(\sqrt{\nu}\right)$ the mixing angle, with $\nu=n/n_{\rm{crit}}$. 
This dressing has consequences both on the qubit and the cavity dynamics. For example, the operator $\sigma_{-}$ responsible for relaxation in the undressed qubit has a reduced effect on the dressed qubit since $|\overline{\bra{g,n}}\sigma_- \overline{\ket{e,n}}|^2 =  c_{n}^2c_{n+1}^2$ is smaller than 1. This is the ``dressed relaxation'' illustrated in Figs.~\ref{rates_renorm}(a) and (a''). But it now also leads to ``relaxation-induced excitation'' since $|\overline{\bra{e,n-2}}\sigma_-\overline{\ket{g,n}}|^2 =  s_{n-1}^2s_n^2$ is non-zero: a dressed ground state can be excited to a dressed excited state through $\sigma_{-},$ a very peculiar effect illustrated in Fig.~\ref{rates_renorm}(d') and (d''). Non-zero matrix elements $|\overline{\bra{e,n-1}}\sigma_- \overline{\ket{e,n}}|^2$ and $|\overline{\bra{g,n-1}}\sigma_- \overline{\ket{g,n}}|^2$ lead to a modified cavity dynamics.

The changes in the matrix elements involving $\sigma_{-,+,z}$, $a$ and $a^{\dagger}$ are derived in Appendix~\ref{app:Derivation}. The transition rates also involve the spectral density of the noise at the frequency characteristic to each process \cite{Boissonneault}. As far as the qubit dynamics is concerned, the results are summarized in Fig.~\ref{rates_renorm}:
\begin{itemize}
    \item The ``dressed relaxation'' rate $\Gamma^{\sigma_-}_\downarrow$(Fig.~\ref{rates_renorm}(a)) is the renormalization of the standard relaxation rate and involves the ability of the bath coupled to the qubit to absorb a photon at frequency $f_q$, given by $S_{\perp}(f_q)$. ``Dressed excitation'' is the reverse process, with a rate $\Gamma^{\sigma_+}_\uparrow$ governed by the bath emission ability given by $S_{\perp}(-f_q)$. The renormalization of both terms, shown in Fig.~\ref{rates_renorm}(a''), follows $|\overline{\bra{g,n}}\sigma_- \overline{\ket{e,n}}|^2=|\overline{\bra{e,n}}\sigma_+ \overline{\ket{g,n}}|^2=c_n^2c_{n+1}^2$ (see  Eqs.~(\ref{EQ:down_sigma-}) and (\ref{EQ:up_sigma+})).
    \item Figure~\ref{rates_renorm}(b,b') deals with processes involving the bath coupled to the cavity. The Purcell relaxation rate $\Gamma^{a}_\downarrow$ is proportional to the ability for the cavity bath to absorb a photon at frequency $f_q$, given by $S_{\kappa}(f_q)$. At finite temperature, the bath can emit photons at frequency $f_q$ and excite the qubit: we call this process ``inverse-Purcell excitation'', its rate is $\Gamma^{a^\dagger}_\uparrow$. The photon-number dependence of both rates shown in Fig.~\ref{rates_renorm}(b'') is given by $|\overline{\bra{g,n}}a^{} \overline{\ket{e,n}}|^2=|\overline{\bra{e,n}}a^{\dagger} \overline{\ket{g,n}}|^2$ (see  Eqs.~(\ref{EQ:down_a}) and (\ref{EQ:up_adag})).
    \item The operator $\sigma_z$, which leads to dephasing of the uncoupled qubit, causes relaxation (rate $\Gamma^{\sigma_z}_\downarrow$) and excitation (rate $\Gamma^{\sigma_z}_\uparrow$) between the dressed states, as illustrated in Fig.~\ref{rates_renorm}(c,c'). The rates are proportional to $\overline{\bra{g,n+1}}\sigma_z \overline{\ket{e,n}}|^2$ and $\overline{\bra{e,n-1}}\sigma_z \overline{\ket{g,n}}|^2$, which coincide when $n\gg 1$ (dependency on $n$ shown in Fig.~\ref{rates_renorm}(c'')) and to the absorption and emission capabilities of the bath coupled to the qubit at $\pm (f_q-f_r)$, associated to the spectral noise density $S_{\parallel}$ (see Eqs.~(\ref{EQ:down_sigmaz}) and (\ref{EQ:up_sigmaz})). 
    \item The transverse noise in the bath coupled to the qubit at $f_q-2f_r$ gives rise to ``excitation-induced relaxation'', in which the operator $\sigma_+$ that causes excitation of the uncoupled qubit allows relaxation  of the dressed qubit, with a 2-photon excitation of the cavity and the emission of one photon at $2f_r-f_q$ by the bath (we assume here that, as it is the case in our experiments, this frequency is positive, otherwise the photons go in the opposite way). The reverse process is ``relaxation-induced excitation''. The rates for both processes $\Gamma^{\sigma_+}_\downarrow$ and $\Gamma^{\sigma_-}_\uparrow$ are proportional to $|\overline{\bra{g,n+2}}\sigma_+ \overline{\ket{e,n}}|^2$ and $|\overline{\bra{e,n-2}}\sigma_- \overline{\ket{g,n}}|^2$, which coincide when $n\gg 1$ (Fig.~\ref{rates_renorm}(d''), see Eqs.~(\ref{EQ:down_sigma+}) and (\ref{EQ:up_sigma-})). 
    \item Finally, there are processes that exchange photons between the cavity and its bath at $2f_r-f_q,$ while photons at $f_q$ go to or come from the qubit. Because they involve the operators $a^{\dagger}$ and $a$ that cause excitation and relaxation of the uncoupled cavity, we call the process illustrated in Fig.~\ref{rates_renorm}(e) ``cavity-excitation-induced relaxation'' and the reverse process ``cavity-relaxation-induced excitation''. Their rates $\Gamma^{a^\dagger}_\downarrow$ and $\Gamma^{a}_\uparrow$ are governed by $|\overline{\bra{e,n-2}}a^{} \overline{\ket{g,n}}|^2$ and $|\overline{\bra{g,n}}a^{\dagger} \overline{\ket{e,n-2}}|^2$ which coincide when $n\gg 1$ (Fig.~\ref{rates_renorm}(e''), see Eqs.~(\ref{EQ:down_adag}) and (\ref{EQ:up_a})).
\end{itemize}

%\paragraph{Renormalized dynamics for a driven cavity ---}
For the driven cavity the system will evolve to a steady-state characterized by a mean number of photons $\overline{n}$ and a density matrix $\hat{\rho}$ from which the probability distribution for the different states $P_{\hat{\rho}}(\overline{\ket{n,g}})$, and $P_{\hat{\rho}}(\overline{\ket{n,e}})$ can be computed. With this in mind, and defining:
\begin{equation}
    \begin{aligned}
        \Gamma^{\text{tot}}_\downarrow(n)&=\left(\Gamma^{\sigma_-}_\downarrow + \Gamma^{a}_\downarrow + \Gamma^{\sigma_z}_\downarrow+\Gamma^{\sigma_+}_\downarrow+\Gamma^{a^{\dagger}}_\downarrow\right)(n)\\
        \Gamma^{\text{tot}}_\uparrow(n)&=\left(\Gamma^{\sigma_+}_\uparrow + \Gamma^{a^{\dagger}}_\uparrow+\Gamma^{\sigma_z}_\uparrow+ \Gamma^{\sigma_-}_\uparrow + \Gamma^{a}_\uparrow \right)(n),
    \end{aligned}
\end{equation}
we have
\begin{equation}
    \begin{aligned}
        \Gamma_{\downarrow}(\overline{n})&= \sum_n P_{\hat{\rho}}(\overline{\ket{n,e}}) \Gamma^{\text{tot}}_\downarrow(n) \\
\Gamma_{\uparrow}(\overline{n})&= \sum_n P_{\hat{\rho}}(\overline{\ket{n,g}})\Gamma^{\text{tot}}_\uparrow(n) \\
\Gamma_{\phi}(\overline{n})&= \sum_n P_{\hat{\rho}}(\overline{\ket{n,e}}) \Gamma^e_\phi(n) + P_{\hat{\rho}}(\overline{\ket{n,g}}) \Gamma^g_\phi(n).
    \end{aligned}
\label{eq_fit_rates}
\end{equation}

We use these expressions to fit the experimental results for the relaxation and excitation rates with two assumptions: i) that the probability distribution corresponds to that of a coherent state, and ii) that the crossed process $\Gamma^{\sigma_+}_\downarrow, \Gamma^{a^{\dagger}}_\downarrow, \Gamma^{\sigma_-}_\uparrow $ and $ \Gamma^{a}_\uparrow $  play only a minor role.  

\subsection{Experimental determination of rates}
In order to obtain information on the dynamics, we analyze simultaneously the one-second-long time-traces $I(t)$ and $Q(t)$ with a hidden Markov model (HMM), using the freely available SMART package \cite{SMART}, yielding the position of the clouds for the states $\ket{g},$ $\ket{e},$ and $\ket{o}$ and the 6 transition rates between them. We do this analysis for the time-traces taken at different powers, for a series of contacts having different Andreev frequencies. The Andreev frequency is measured by two-tone spectroscopy before and after taking the data as a function of power, to ensure that the atomic contact remains the same within our experimental uncertainty.

\begin{figure}[t!]
        \includegraphics[width=\columnwidth]{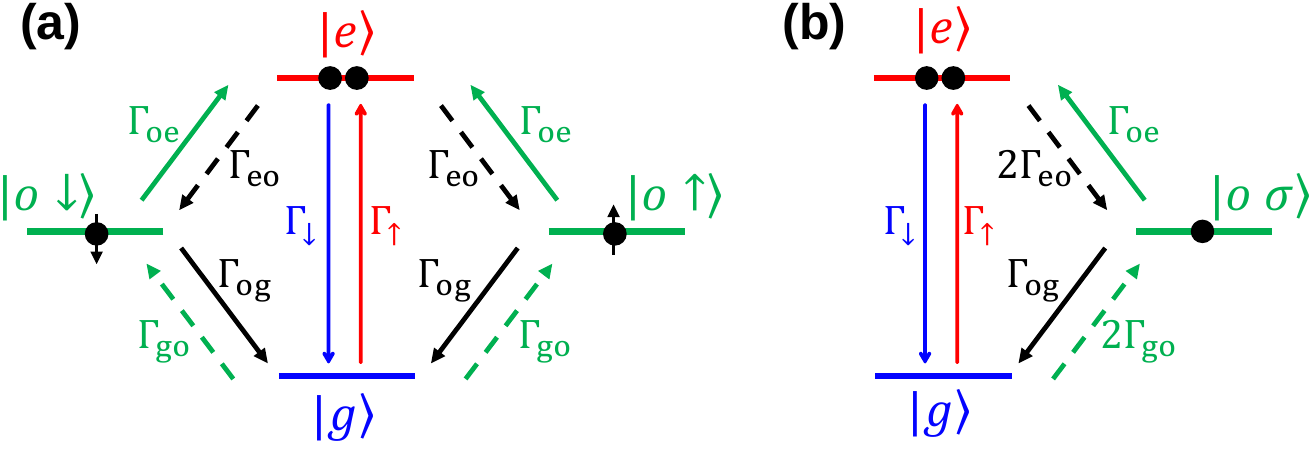}
\caption{\textbf{Definition of the transition rates between states.} (a) Rates between all states, with the distinction between odd states with different spin $|o_\downarrow\rangle$ and $|o_\uparrow\rangle$. 
(b) Equivalent diagram with a spin-degenerate odd state.}
\label{rates_btw_states}
\end{figure}

The transitions rates within the even manifold $\ket{g} \leftrightarrow \ket{e}$ are the Andreev qubit excitation and relaxation rates $\Gamma_{\uparrow}$ and $\Gamma_\downarrow$. The remaining four rates correspond to parity jumps. As illustrated in Fig.~\ref{rates_btw_states}(a), we note $\Gamma_{\rm oe}$ and $\Gamma_{\rm og}$ the rates of the processes that add or remove a quasiparticle in the Andreev level, starting from an odd state of either spin ($\ket{o_\uparrow}$ or $\ket{o_\downarrow}$), $\Gamma_{\rm go}$ the rate for adding a quasiparticle starting from $\ket{g}$, and $\Gamma_{\rm eo}$ that for removing a quasiparticle starting from $\ket{e}$. The system being spin-degenerate, these rates do not depend on spin. The relation between these definitions and the rates extracted from the time-trace fitting is shown in Fig.~\ref{rates_btw_states}(b), using a representation in which no distinction is made between the two odd states $\ket{o_\uparrow}$ and $\ket{o_\downarrow}$. The 6 rates inferred from the analysis of the traces are therefore identified to $\Gamma_\downarrow,$ $\Gamma_{\uparrow},$ $2\Gamma_{\rm eo},$ $\Gamma_{\rm oe},$ $\Gamma_{\rm og},$ and $2\Gamma_{\rm go}$ (note the factors 2 \cite{Catelani}). 

\begin{figure*}[t!]
\includegraphics[width=2.1\columnwidth]{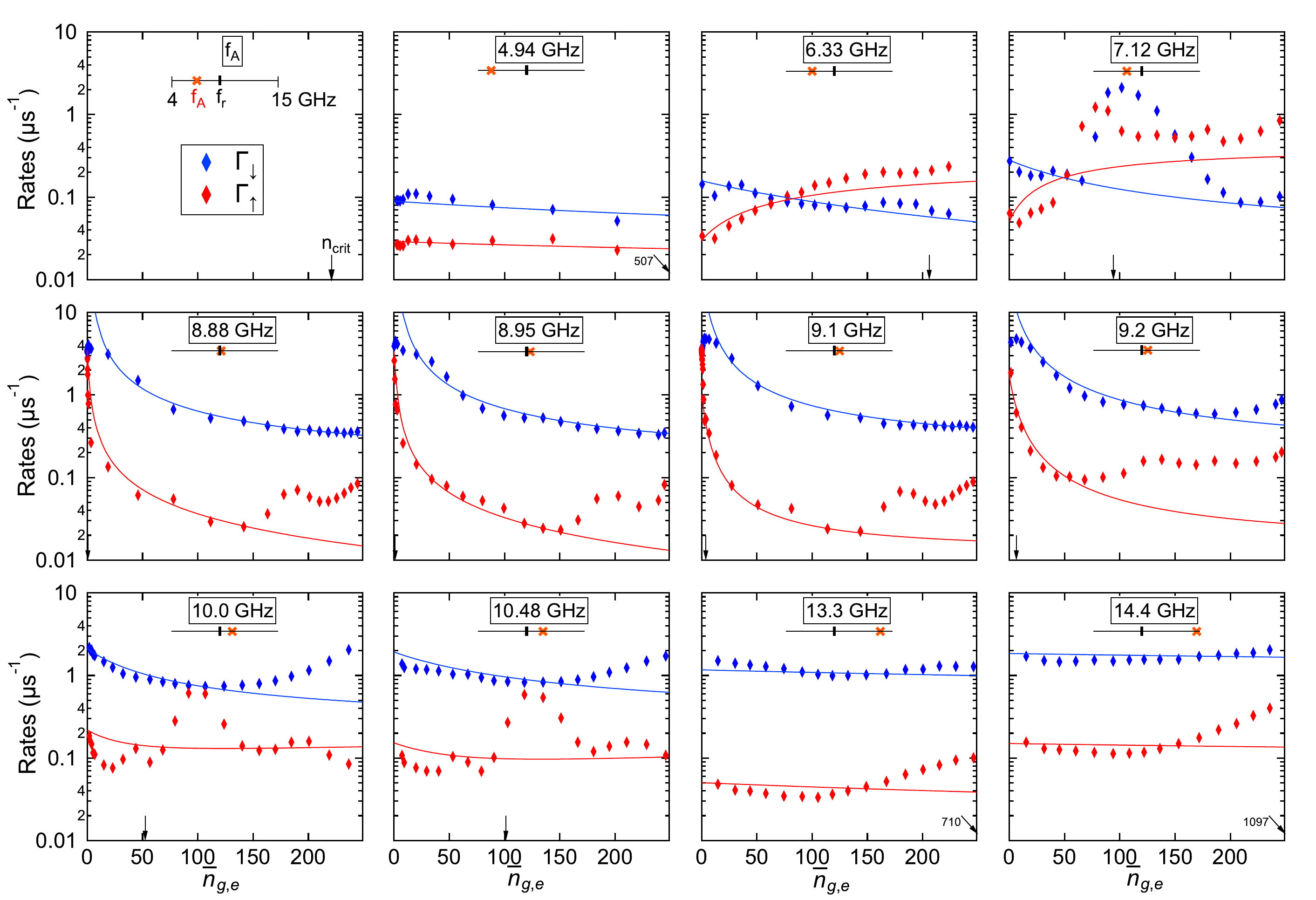}
\caption{\textbf{Relaxation and excitation transition rates as a function of photon number} for different contacts ($f_A$ indicated on each panel from $f_A=4.94$~GHz to 14.4~GHz, with symbolic representation of the relative position of $f_A$ (red cross) relatively to $f_r$ (black tick) on a segment representing the interval 4--15~GHz). Arrows on the x-axis indicate the value of $n_{\rm crit}$. Rates are obtained from the analysis of time traces with a boxcar average of 10 points.
Continuous lines correspond to calculated dependencies using the theoretical expressions in Eq.~(\ref{eq_fit_rates}) with prefactors shown in Fig.~\ref{ratesfreqfitcoeff}.}
\label{powerdepeven}
\end{figure*}

\begin{figure*}
\includegraphics[width=2.1\columnwidth]{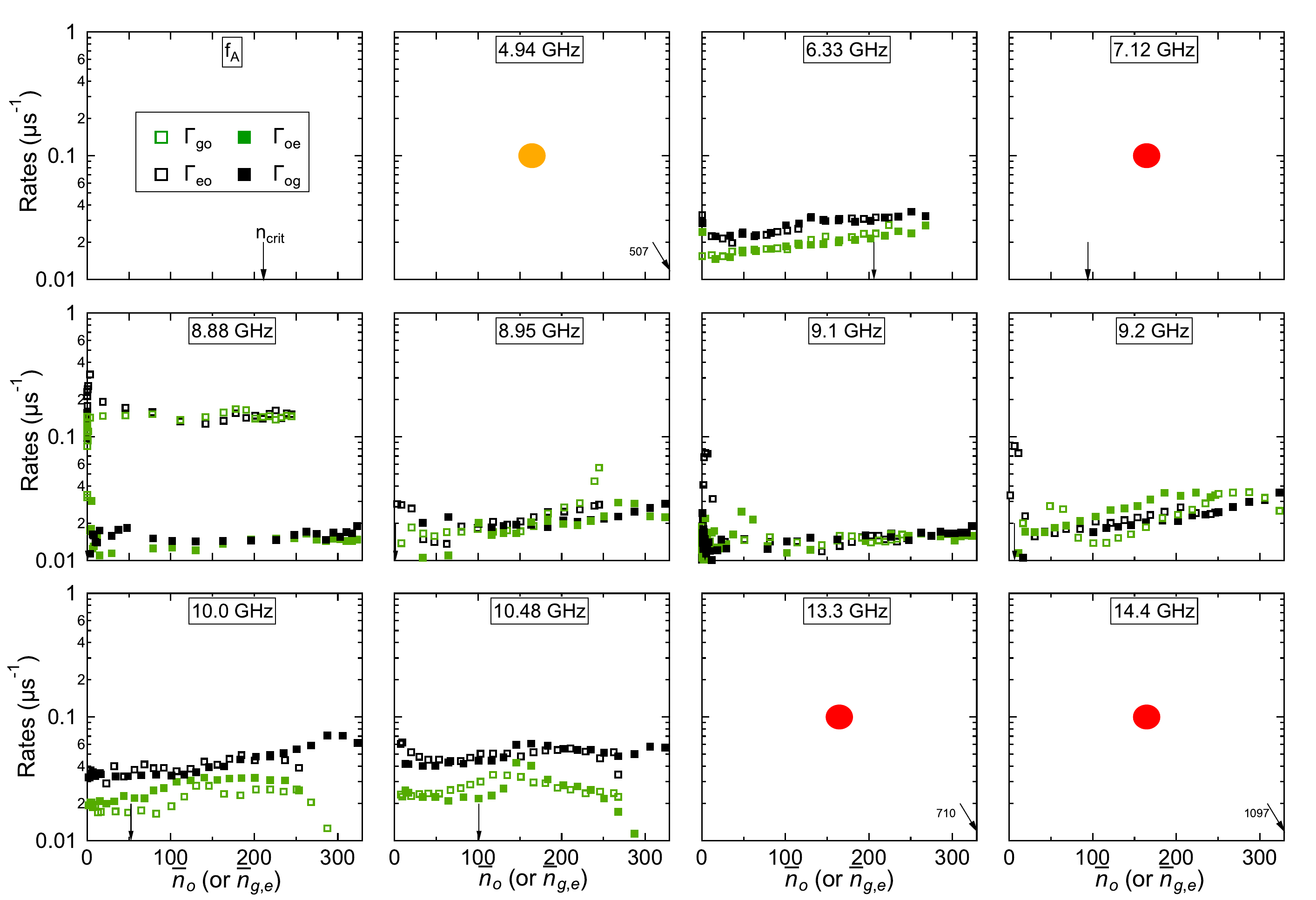}
\caption{\textbf{Parity jump transition rates as a function of photon number} for different contacts ($f_A$ indicated on each panel from $f_A=4.94$~GHz top-left to 14.4~GHz right-bottom). Rates are obtained from the analysis of time traces with a boxcar average of 10 points. Orange and red disks signal results that depend significantly on filtering, and hence are less significant (see Appendix~\ref{app:timetrace}).}
\label{powerdepodd}
\end{figure*}

\subsection{Rates vs. photon number}
In  Figs.~\ref{powerdepeven} and \ref{powerdepodd} we show the transition rates as a function of the photon number for different Andreev frequencies. From top-left to bottom-right the Andreev qubit frequency changes from 4.94~GHz to 14.4~GHz, crossing the cavity frequency at 8.77~GHz. Figure~\ref{powerdepeven} shows the rates between $\ket{g}$ and $\ket{e}$ (quantum jumps), whereas Fig.~\ref{powerdepodd} shows the rates from or to $\ket{o}$ (parity jumps).  The control parameter is the power of the continuous probe tone at $f_r,$ translated into the average photon number $\overline{n}_g,$ $\overline{n}_e$ or $\overline{n}_o$ in the initial state using the calibration described in Section \ref{DDROP}. The arrows on the x-axes indicates the value of $n_{\text{crit}}$ for each case. Note that since we are using a drive tone at $f_r,$ $\overline{n}_g=\overline{n}_e$ if one neglects the small cavity pull associated to other, less-coupled Andreev states. In addition, $\overline{n}_{g,e}$ is smaller than $\overline{n}_o$ because $f_r$ is the resonance frequency of the cavity when the system is in $\ket{o}$. Finally, it is worth mentioning that $\overline{n}_{g,e,o}$ recover their steady state value within $2/\kappa \approx 35~$ns when a parity change occurs, a time short enough to consider that all the rates correspond to their value for the steady state photon number.

For each time trace, we performed the analysis with different box-car filters (see Appendix \ref{app:timetrace}). Data filtering increases the signal-to-noise (SNR) ratio and facilitates states discrimination, but this comes at the price of rounding the transitions and averaging out fast back-and-forth transitions. As a consequence, the rates inferred from the analysis sometimes depend on the filtering, but the less-filtered data are eventually unreliable because the SNR is too small. 
This is discussed at length in Appendix  \ref{app:timetrace}. In Fig.~\ref{powerdepeven} we show results for datasets in which the quantum jump rates are essentially independent of the smoothing. Figure~\ref{powerdepodd} shows the parity jump rates for the same datasets. For four values of $f_A$, marked with orange or red color disks, the rates inferred from the analysis are unreliable because they depend strongly on the smoothing.
%~\ref{powerdepeven} and \ref{powerdepodd}. We have data sets where the rates depend strongly on the filtering with orange or red warning disks at the top right of the panels.}

\subsubsection{Quantum jumps}
As can be observed in Fig.~\ref{powerdepeven}, $\Gamma_{\uparrow}$ (red symbols) and $\Gamma_{\downarrow}$ (blue symbols) depend both on photon number and on $f_A$. Consider first the dependence of the relaxation rate at vanishing photon number $\Gamma_{\downarrow}(0)$.
As $f_A$ approaches $f_r$, $\Gamma_{\downarrow}(0)$ increases considerably, indicating that the emission of photons to the cavity plays an important role. The excitation rate at $\overline{n}_g=0$ shows a similar increase close to $f_r,$ which indicates that the cavity temperature is sufficiently high to excite the Andreev states in a reverse process. In addition, there is no symmetry between negative and positive detuning: $\Gamma_{\downarrow}(0)$ is much larger when $f_A>f_r$ than at $f_A<f_r$. 

As far as the photon number dependence is concerned, one observes for several contacts a strong increase of the rates at intermediate (for $f_A=7.12$, 7.47 and 8.2~GHz) or large ($f_A=9.2,$ 10 and 10.48~GHz) photon number. This cannot be reproduced by the theory of Section \ref{Rates renormalization}, except perhaps if one assumed the existence of broad environmental modes at some 
frequencies. We suspect that transitions involving other Andreev states or an effect of the JPC pump tone could play a role. 
In the following discussion, we focus on what happens at smaller photon numbers. The overall tendency is that $\Gamma_{\downarrow}$ decreases with $\overline{n}$, particularly when the detuning with the cavity is smaller. No systematic is observed for $\Gamma_{\uparrow}$. 
Solid lines correspond to theoretical dependencies following the rates renormalization theory presented in Section \ref{Rates renormalization}. In this comparison, we neglected the slight dependence of $g$ with $f_A$. The prefactors for the various contributions to the rates were adjusted to account at best for the overall dependencies, in particular at low photon number.
The rapid decay of the relaxation rate at low $\overline{n}_e$ and its increase at small detuning reveal the dominant contribution of the Purcell effect. 
An additional contribution of dressed relaxation is needed to account for the variations at large $\overline{n}_e$. We found that the other terms (dephasing-induced, excitation-induced or cavity-excitation-induced relaxation) have no significant contribution.

As for the excitation rate, one needs to consider the thermal photons in the cavity that, like in Purcell effect, account for its decaying behavior with $\overline{n}_e$ when $f_A \approx f_r.$ The opposite behaviour observed far from $f_r$ when $f_A < f_r $ (see for example $f_A=6.33$ and $7.12$~GHz) reveals a contribution of the dressed dephasing $\Gamma_{\uparrow}^{\sigma_z}.$ 

\subsubsection{Parity jumps}
Parity jump rates are shown in Fig.~\ref{powerdepodd}. For many contacts, signaled with orange or red disks, the rates extracted from the data depend strongly or very strongly on filtering. These data are therefore inconclusive, and we focus here on the others.
A first observation is that the rates corresponding to the addition (resp., the removal) of a quasiparticle in the Andreev level do not depend on the level occupancy: $\Gamma_{go}\approx \Gamma_{oe}$ (resp., $\Gamma_{eo}\approx \Gamma_{og}$). This is expected in a short weak link, in which charging effects do not play a significant role \cite{Matute2022,Fatemi2022}. A second observation is that the rates $\Gamma_{go}$ and $\Gamma_{oe}$ (addition of a quasiparticle, green symbols) are systematically smaller than $\Gamma_{eo}$ and $ \Gamma_{og}$ (removal of a quasiparticle, black symbols). This is discussed in the next section. 
The only contact for which these two observations are not obeyed is the one with the smallest detuning, at $f_A=8.88~$GHz.

At low photon number, the rates are mostly in the 0.01-0.1~$\mu$s$^{-1}$ range, corresponding to parity lifetimes from 10 to 100~$\mu$s. This is slightly smaller or comparable with results from other experiments on atomic contacts \cite{Zgirski, Janvier} and on nanowire weak links \cite{Hays2018, Hays2020}. The dependence of the rates on the number of photons is less pronounced than for the relaxation and excitation rates. In most contacts, there is no significant dependence at all. From the theoretical point of view, we are not aware of any work addressing this issue.

\section{Intrinsic dynamics: Implications for Andreev physics}

\begin{figure}
\includegraphics[width=\columnwidth]{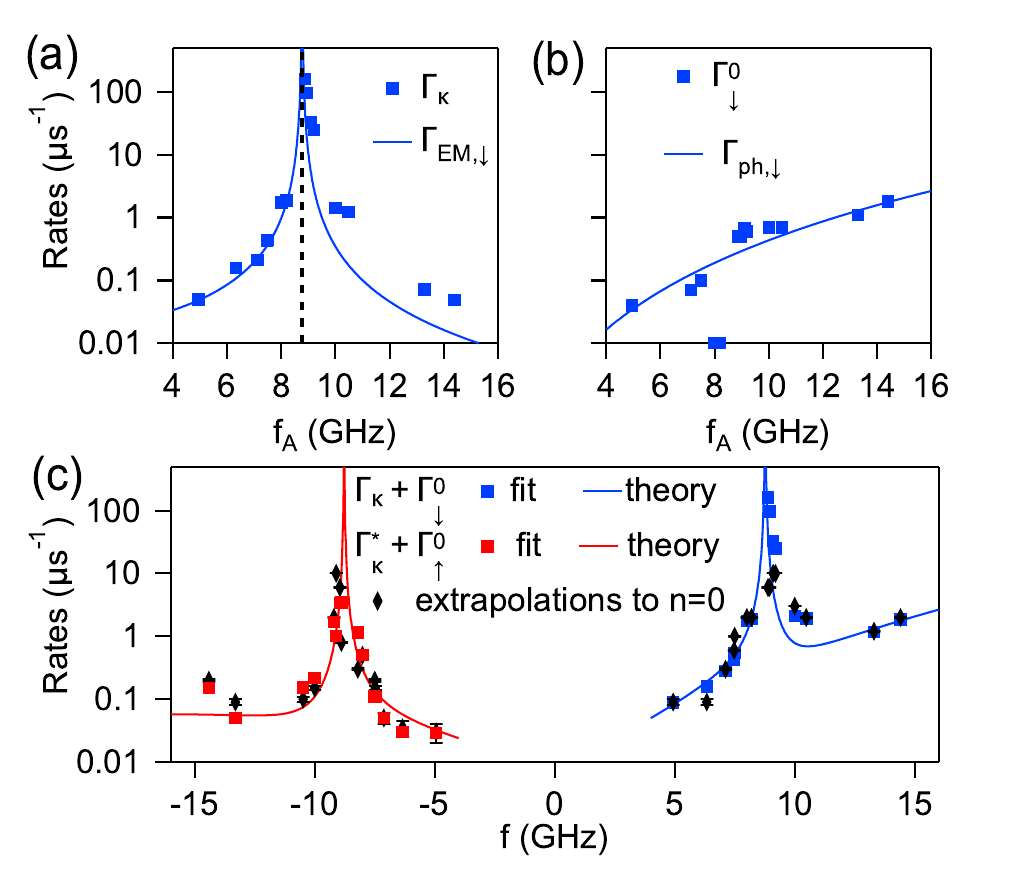}
\caption{\textbf{Intrinsic relaxation and excitation rates as a function of Andreev frequency $f_A$.}
(a) Comparison of the Purcell rate $\Gamma_\kappa=\kappa^0[f_A]\left(\frac{g}{\Delta}\right)^2$ extracted from the fits (symbols) with theoretical photon emission rate $\Gamma_{\rm EM, \downarrow}$ (solid line), using $T_{\text{EM}}=300$~mK. 
(b) Intrinsic relaxation rate $\Gamma_\downarrow^0$ (symbols) compared with phonon emission rate $\Gamma_{\rm ph,\downarrow}$ (solid lines), using  
$C_{\rm ph}=40~{\rm s}^{-1}{\rm GHz}^{-4}$ and $T_{\text{ph}}=200$~mK. 
(c) Total intrinsic relaxation rate $\Gamma_\kappa+\Gamma_\downarrow^0$ (red squares) and emission rate $\Gamma_\kappa^*+\Gamma_{\uparrow}^0$ (blue squares), from the analysis of the full dependence of the rates with photon number. Rates from the extrapolation of the data at $\overline{n}=0$ (black diamonds). Solid lines are comparison with theory describing photon and phonon emission and absorption.}
\label{ratesfreqfit}
\end{figure}

\subsection{Andreev qubit}
We discuss here the rates found in the zero-photon limit. They are extracted either by extrapolation of the data to $\overline{n}=0$, or by taking the $n=0$ limit of the theory curves that are adjusted to the data. The relaxation and excitation rates simplify at $n=0$ to:
\begin{equation}
    \begin{aligned}
        \Gamma_{\downarrow}(0)&= \Gamma^0_\downarrow+ \Gamma_\kappa, \\
        \Gamma_{\uparrow}(0)&= \Gamma^0_\uparrow+\Gamma_{\kappa}^{*},
    \end{aligned}
\end{equation}
with $\Gamma_{\kappa}=\kappa^{0}[f_A]\left(\frac{g}{\Delta}\right)^2$ the Purcell rate and $\Gamma_{\kappa}^{*}\equiv\Gamma_{\uparrow}^{a^\dagger}(0)=\kappa^{0}[-f_A]\left(\frac{g}{\Delta}\right)^2$ the inverse process in which thermal photons in the cavity excite the qubit. The rates $\Gamma^0_\downarrow$ and $\Gamma^0_\uparrow$ as well as the Purcell prefactors $\kappa^{0}[f_A]$ and $\kappa^{0}[-f_A]$ used to adjust the data are shown in Fig.~\ref{ratesfreqfitcoeff}. The corresponding dependence of the rates are shown with symbols in Fig.~\ref{ratesfreqfit}. The function $\kappa^{0}[f]$ describes the ability of the cavity bath to absorb photons at frequency $f$. In particular, $\kappa^{0}[f_r]=\kappa.$

The rate $\Gamma_\downarrow^0$ is expected to correspond to the emission of phonons, whereas the Purcell rate accounts for the emission of photons. At zero temperature, the phonon emission rate reads, for transmissions $\tau$ close to 1 \cite{Olivares, Zazunov2005, thesisJanvier,Anthore}:
\begin{equation}
    \Gamma_{\text{ph}}=\kappa_{\rm ph} \frac{\Delta_s (1-\tau)}{3 E_A} E_A^3,
\end{equation}
with $\kappa_{\rm ph}$ the  electron-phonon coupling constant \cite{Anthore}. At $\varphi=\pi,$ $E_A=\Delta_s \sqrt{1-\tau}$ and the expression simplifies to:
\begin{equation}
    \Gamma_{\text{ph}}=\kappa_{\rm ph} \frac{E_A }{3\Delta_s} E_A^3,
\end{equation}
and keeping only the Andreev frequency dependence:
\begin{equation}
    \Gamma_{\text{ph}}=C_{\text{ph}} f_A^4,
\end{equation}
where $C_{\text{ph}}$ is a constant that depends on the electron-phonon coupling in the aluminium constriction.
At finite temperature, phonons contribute both to relaxation ($\Gamma_\downarrow^0$) and excitation ($\Gamma_{\uparrow}^0$):
\begin{equation}
    \begin{aligned}
        \Gamma_{\text{ph},\downarrow}&=\Gamma_{\text{ph}} \times \left(1+n_{\text{ph}}\right)\\
        \Gamma_{\text{ph},\uparrow}&=\Gamma_{\text{ph}} \times n_{\text{ph}},
    \end{aligned}
\end{equation}
with $n_{\text{ph}}$ the Bose population factor at the phonon temperature $T_{\text{ph}}$. 

As far as photons are concerned, a precise prediction of the Purcell effect must take into account the full frequency dependence of the impedance seen by the atomic contact \cite{Desposito}. Approximating the $\lambda/4$ CPW resonator with a single mode cavity, one obtains, at $\varphi=\pi,$ and assuming $\Delta \gg \frac{\kappa}{2}$ \cite{thesisJanvier}:
\begin{equation}
    \Gamma_{\text{EM}}=\kappa\left(\frac{g}{\Delta}\right)^2 \frac{1+\frac{\Delta}{\omega_r}}{\left(1+\frac{\Delta}{2\omega_r}\right)^2},
\end{equation}
leading, at finite temperature, to contributions to both relaxation ($\Gamma_\kappa$) and excitation ($\Gamma_\kappa^*$):
\begin{equation}
    \begin{aligned}
        \Gamma_{\text{EM},\downarrow}&=\Gamma_{\text{EM}}\times\left(1+n_{\text{EM}}\right)\\
        \Gamma_{\text{EM},\uparrow}&=\Gamma_{\text{EM}} \times n_{\text{EM}},
    \end{aligned}
\end{equation}
where $n_{\text{EM}}$ is the Bose function corresponding to the temperature of the electromagnetic environment $T_{\text{EM}}$.

Figure~\ref{ratesfreqfit} shows comparisons between experiment and theory. The fit temperatures are significantly larger than that of the mixing chamber in the experiment. When varying the temperature of the experiment, we observed that the rates only started to change above $\SI{200}{\milli\K},$ indicating that the sample environment was indeed hot. Overall, we find that the intrinsic rates are well described by effects of phonons and cavity photons. However, the amplitude of the prefactor for the electron-phonon interaction $C_{\rm ph}$ is $\approx20$ times larger than measurements in aluminum wires \cite{Anthore,Reulet}. This discrepancy might be related to the geometry of the atomic-size contact.

\begin{figure}[t!]
\includegraphics[width=1\columnwidth]{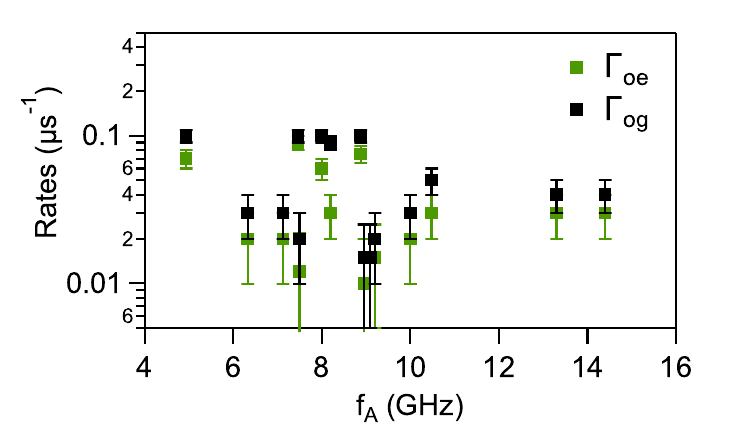}
\caption{\textbf{Intrinsic parity jumps transition rates as a function of $f_A$} obtained from extrapolation of data of Fig.~\ref{powerdepodd} towards $\overline{n}=0$.}
\label{ratesfreqfitodd}
\end{figure}

\subsection{Parity switching in atomic contacts}
The parity jump rates $\Gamma_{go}$, $\Gamma_{oe}$, $\Gamma_{eo}$ and $\Gamma_{og}$ correspond to processes that involve quasiparticles in the continuum with an energy $E_{\rm{qp}}$ larger than the superconducting gap $\Delta_s$ \cite{Zgirski,Kos, Olivares}. The transition $\ket{g} \rightarrow \ket{o}$ corresponds to a quasiparticle at $E_{\rm{qp}}$ that relaxes into the Andreev level, with emission of a photon or phonon at $E_{\rm{qp}}-E_A.$ The reverse process involves the recombination of the quasiparticle at $E_{\rm{qp}}$ with that in the Andreev level, and emission of a photon or a phonon at $E_{\rm{qp}}+E_A.$ Altogether, one predicts 
\begin{align}
\Gamma_{go,oe}&\propto f(E_{\rm{qp}}) D(E_{\rm{qp}}-E_A),\nonumber\\
\Gamma_{eo,og}&\propto f(E_{\rm{qp}}) D(E_{\rm{qp}}+E_A),
\end{align}
with $D(E)$ the density of modes in the environment and $f(E_{\rm{qp}})$ the occupation factor of the quasiparticle state at $E_{\rm{qp}}.$ The observation that $\Gamma_{go},\Gamma_{oe} < \Gamma_{eo},\Gamma_{og}$ corresponds to $D(E)$ being an increasing function of $E,$ which is expected for phonons. The number of photons in the cavity plays a minor role in these processes. The fluctuations of the rates from one measurement to another indicate that the density of quasiparticles in the continuum varies at time scales of hours or days, in an uncontrolled manner, see Fig.~\ref{ratesfreqfitodd}. This can be related to the fluctuations of the characteristic times of superconducting qubits.

\section{Discussion and Conclusions}
We have shown how superconducting atomic contacts allow probing the physics of  cQED when varying the photon number $n$ in the cavity. With contacts detuned from the cavity by 5.63~GHz to 0.11~GHz, we could explore situations with $n_{\rm crit}$ from 0.4 to 1140, while $n$ was varied from 0 to 250. The evolution of the position of the clouds corresponding to the states of the system illustrate how the cavity is rendered non-linear by its coupling to the Andreev qubit. This effect is also well seen in experiments in which a 2-tone spectroscopy is performed in presence of an additional tone at a frequency close to that of the cavity (DDROP measurements). We observe strong changes in the transition rates between the states of the Andreev qubit. We have extended the existing theories, which were limited to $n \ll n_{\rm crit}$, to account for the data. A systematic analysis of how the different operators that affect the bare qubit change the rates between the dressed states reveals a great variety of processes, which involve the baths coupled to the qubit and to the cavity at various frequencies (see Fig.~\ref{rates_renorm}). In the experiment, the rates are inferred from 1-s-long measurements of the quadratures $I(t)$ and $Q(t)$ in presence of photons in the cavity. By investigating the effect of data filtering, we found that this analysis does not always allow extracting reliable information. Still, general effects predicted by theory could be recognized, essentially the dressed Purcell effect and its thermal counterpart, dressed relaxation and dressed dephasing. The rates at low $n$ are well accounted for by the combination of emission and absorption of photons (Purcell effect) and of phonons, although the amplitude of this last term is found more that one order of magnitude larger than in wires. No theory predicts how parity jumps should depend of the photon number; the experiments show that this dependence is weak. A  conclusion of this work is that, when measuring a strongly anharmonic qubit, like the Andreev qubit, the scale for the modification of its dynamics when coupled to a cavity is the photon number $n_{\rm crit}$. In a continuous measurement meant at extracting the transition rates, one can ensure that $n \ll n_{\rm crit}$ when the positions of the clouds in the $(I,Q)$ plane change linearly with measurement power. In a pulsed measurement, the measurement time being set much smaller than the inverse of the highest transition rate, increasing the power only becomes detrimental when the rates of the processes shown in Fig.~\ref{rates_renorm}(c,d,e) start to be significant. The others, which generally dominate, decay with power if the noise spectrum does not present any singularity.
\appendix
\renewcommand{\thefigure}{\thesection.\arabic{figure}}

\section{Clouds positions in $(I,Q)$ plane, additional data}
\label{app:clouds}
\setcounter{figure}{0}
\begin{figure}[h!]
\includegraphics[width=\columnwidth]{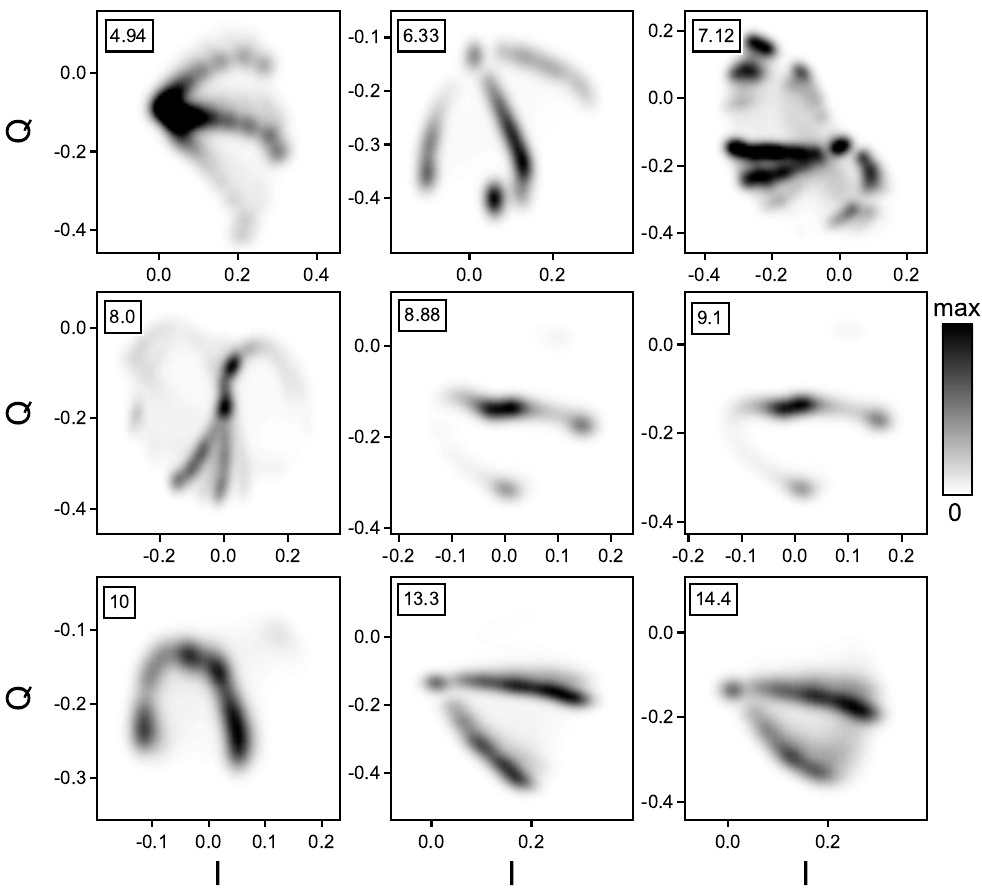}
\caption{Histograms of all the values of $(I,Q)$ recorded during continuous measurements at various probe amplitudes for the same 9 contacts as in Fig.~\ref{histos}, with $f_A$ given in GHz in the text box.}
\label{histos2}
\end{figure}
Figure~\ref{histos2} shows the evolution with measurement amplitude, for the same 9 contacts as in Fig.~\ref{histos}. The data at different amplitudes were concatenated and histogrammed. At small $f_A,$ three branches, corresponding to states $\ket{g}$, $\ket{o}$ and $\ket{e}$ are clearly visible. For the contact at $f_A=7.12~$GHz, they split, each into three, indicating the presence of a second Andreev level with a larger detuning and found either in its ground, odd or excited state. This is also the case at $f_A=8.0~$GHz, but less visible. The clouds corresponding to the excited state are almost invisible in the data for the contacts at larger $f_A$, the steady state occupation of $\ket{e}$ becoming negligible compared to $\ket{g}$ and $\ket{o}$.
\begin{table}[h!]
\centering
 \begin{tabular}{||c c c c c||} 
 \hline
 $f_A$ (GHz) & $n_{\rm crit}$ & $\chi_0/\kappa$ & $\delta \omega/\kappa$ & values of $\sqrt{n_0}$ in Fig.~\ref{histos}\\ [0.5ex] 
 \hline\hline
 4.94 & 504 & -0.20 & 0 & 1.7, 6.1, 10, 13, 15\\ 
 6.33 & 205 & -0.32 & 0 & 0.8, 4.1, 8.1, 12, 16\\
 7.12 & 94 & -0.47 & 0.15 & 0.8, 4.1, 8.1, 12, 16\\
 8.00 & 21 & -1.0 & 0.043 & 0.8, 4.1, 8.1, 12, 16\\
 8.88 & 0.42 & 7.1 & 0 & 0.8, 3.8, 7.8, 13, 17\\  
 9.10 & 3.8 & 2.4 & 0 & 0.8, 3.8, 7.8, 13, 17\\
 10.0 & 53 & 0.64 & 0 & 0.8, 4.1, 8.1, 12, 16\\
 13.3 & 734 & 0.17 & 0.12 & 0.8, 4.1, 8.1, 12, 16\\
 14.4 & 1140 & 0.14 & 0.11 & 0.8, 4.1, 8.1, 12, 16\\
 \hline
 \end{tabular}
 
 \caption{Characteristics of the nine contacts shown in Figs.~\ref{histos} and \ref{histos2}: Andreev frequency $f_A,$ critical number of photons $n_{\rm crit},$ ratio of the dispersive shift at vanishing number of photons $\chi_0$ and cavity inverse linewidth $\kappa$, additional  shift $\delta \omega$ added in the comparison with theory in Fig.~\ref{histos}. Last column are the values of $\sqrt{n_0}$ in Fig.~\ref{histos}, corresponding to the five sets of clouds and to the five circles.}
\end{table}

\section{Derivation of the qubit rates modifications with cavity photon number}
\label{app:Derivation}
We derive here the expressions for the modified transition rates that result from the dressing of the qubit with the cavity as described by Eq.~(\ref{dressedstates}). The coefficients of the dressed states can be written as:
\begin{equation}
    c_{n}^2=\frac{1}{2}\left(1+\frac{1}{\sqrt{1+\nu}}\right) \text{ and }
    s_{n}^2=\frac{1}{2}\left(1-\frac{1}{\sqrt{1+\nu}}\right).
\end{equation}
We will also make use of the following functions, which have simple expressions in the dispersive limit valid when $\nu \ll 1$:
\begin{equation}
    \begin{aligned}
        R_{cc}(n)&=c_{n}^4\underset{\nu \ll 1}{\approx} 1-\frac{\nu}{2 }\\
        R_{ss}(n)&=s_{n}^4\underset{\nu \ll 1}{\approx} \frac{\nu^2}{16},\\
        R_{cs}(n)&=4c_{n}^2s_{n}^2=1-\frac{1}{{1+\nu}}\underset{\nu \ll 1}{\approx} \nu,\\
        R_{-}(n)&=(c_{n}^2-s_{n}^2)^2=\frac{1}{1+\nu}
        \underset{\nu \ll 1}{\approx} 1-\nu.
    \end{aligned}
\end{equation}

\subsection{Effects of $\sigma_{-}$}
\label{PAR:sigma-}
From Eqs. (\ref{dressedstates}), one obtains four types of non-zero matrix elements involving $\sigma_{-}$:
\begin{equation}
    \begin{aligned}
        |\overline{\bra{g,n}}\sigma_- \overline{\ket{e,n}}|^2 &=  c_{n}^2c_{n+1}^2,\\
        |\overline{\bra{e,n-1}}\sigma_- \overline{\ket{e,n}}|^2 &= s_{n}^2c_{n+1}^2,\\
        |\overline{\bra{g,n-1}}\sigma_- \overline{\ket{g,n}}|^2 &= c_{n-1}^2s_n^2,\\
        |\overline{\bra{e,n-2}}\sigma_- \overline{\ket{g,n}}|^2 &=  s_{n-1}^2s_n^2.
    \end{aligned}
\end{equation}
The first one leads to a renormalization of the relaxation rate $\Gamma^{\sigma_-}_\downarrow(n)$. We call it ``dressed relaxation''. The second and third one introduce contributions to the photon loss rates. The last one, which is the smallest one when $n \ll n_{\rm crit}$ since it involves a product of two sine coefficients, is more peculiar: it leads to an \textit{excitation} of the qubit caused by the collapse operator $\sigma_{-}$, involving the absorption of two photons. We note the corresponding rate $\Gamma^{\sigma_-}_\uparrow(n),$ and call it ``relaxation-induced excitation''. We illustrate these processes in Figs.~\ref{rates_renorm}(a) and (d).

In addition to the modification of the matrix elements, the complete description of the dressed states dynamics involves the spectral density of the noise at the frequency characteristic to each process \cite{Boissonneault}. We note this frequency as a parameter of each rate, with brackets $[...]$, and obtain:
\begin{equation}
    \begin{aligned}
    \label{EQ:down_sigma-}
        \Gamma^{\sigma_-}_\downarrow(n)&=\Gamma_{\downarrow}^0[\omega_q] c_{n}^2c_{n+1}^2 \\&\underset{\nu \gg 1}{\approx} \Gamma_{\downarrow}^0[\omega_q] R_{cc}(n), \\
    \end{aligned}
\end{equation}
\begin{equation}
    \begin{aligned}
    \label{EQ:up_sigma-}
        \Gamma^{\sigma_-}_\uparrow(n) &= \Gamma^{0}_\downarrow[-\omega_q+2\omega_r] s_{n-1}^2s_{n}^2 \\
        &\underset{\nu \gg 1}{\approx} \Gamma^{0}_\downarrow[-\omega_q+2\omega_r] R_{ss}(n),
    \end{aligned}
\end{equation}
with $\Gamma_{\downarrow}^0[\omega]=\Gamma_{\downarrow}^0[\omega_q] \times \left(S_\perp(\omega)/S_\perp(\omega_q)\right)$.

In the dispersive limit $n\ll n_{\rm{crit}}$, one recovers the result derived by Boissoneault {\it et al.} \cite{Boissonneault2010}:
\begin{equation}
\Gamma^{\sigma_-}_\downarrow(n) \approx \Gamma^{0}_\downarrow[\omega_q] \left(1- \frac{2n+1}{4n_{\rm{crit}}}\right).
\end{equation}
The excitation term, which is much smaller, goes as
\begin{equation}
\Gamma_{\uparrow}^{\sigma_-}(n) \approx \Gamma^{0}_\downarrow[-\omega_q+2\omega_r] \frac{n(n-1)}{16n_{\rm{crit}}^2}.
\end{equation}

\subsection{Effects of $\sigma_+$}
\label{PAR:sigma+}
In the same way, there are four non-zero matrix elements arising from the action of $\sigma_+$ on the dressed states: 
\begin{equation}
    \begin{aligned}
        |\overline{\bra{e,n}}\sigma_+ \overline{\ket{g,n}}|^2 &= c_{n+1}^2c_{n}^2,\\
        |\overline{\bra{g,n+1}}\sigma_+ \overline{\ket{g,n}}|^2 &=  s_{n+1}^2c_{n}^2\\
        |\overline{\bra{e,n+1}}\sigma_+ \overline{\ket{e,n}}|^2 &=  c_{n+2}^2s_{n+1}^2,\\
        |\overline{\bra{g,n+2}}\sigma_+ \overline{\ket{e,n}}|^2 &= s_{n+2}^2s_{n+1} ^2.
    \end{aligned}
\end{equation}
The first one describes the renormalization of the excitation rate $\Gamma^{\sigma_+}_\uparrow(n)$, by the same factor as the relaxation rate. We call this term ``dressed excitation''. The second and third ones are effective contributions to the cavity drive. The last one is a peculiar contribution that leads to the relaxation rate $\Gamma^{\sigma_+}_\downarrow(n)$ (see Fig.~\ref{rates_renorm}(d)). We call it ``excitation-induced relaxation''. One obtains for the two rates corresponding to changes in the qubit state:
\begin{equation}
    \begin{aligned}
    \label{EQ:up_sigma+}
        \Gamma^{\sigma_+}_\uparrow(n) &= \Gamma^{0}_\uparrow[-\omega_q] c_{n+1}^2c_{n}^2 \\
        &\underset{\nu \gg 1}{\approx} \Gamma^{0}_\uparrow[-\omega_q] R_{cc}(n),\\
    \end{aligned}
\end{equation}
\begin{equation}
    \begin{aligned}
    \label{EQ:down_sigma+}
        \Gamma^{\sigma_+}_\downarrow(n) &= \Gamma^{0}_\uparrow[\omega_q-2\omega_r] s_{n+2}^2s_{n+1}^2 \\
        &\underset{\nu \gg 1}{\approx} \Gamma^{0}_\uparrow[\omega_q-2\omega_r] R_{ss}(n).
    \end{aligned}
\end{equation}

\subsection{Purcell renormalization, effect of $a^{}$ and $a^{\dagger}$}
The cavity annihilation operator $a$ has various effects on the dressed system. They are found from the non-zero matrix elements resulting from the action of $a^{}$:
\begin{equation}
    \begin{aligned}
|\overline{\bra{g,n-1}} a^{} \overline{\ket{g,n}}|^2 &=  (\sqrt{n}c_nc_{n-1}+\sqrt{n-1}s_ns_{n+1})^2,\\
|\overline{\bra{e,n-1}} a^{} \overline{\ket{e,n}}|^2 &=  (\sqrt{n}c_nc_{n+1}+\sqrt{n+1}s_ns_{n+1})^2,\\
|\overline{\bra{g,n}}a^{} \overline{\ket{e,n}}|^2 &=  (\sqrt{n}c_{n+1}s_{n}-\sqrt{n+1}s_{n+1}c_{n})^2,\\
|\overline{\bra{e,n-2}}a^{} \overline{\ket{g,n}}|^2 &=  (\sqrt{n}c_ns_{n-1}-\sqrt{n-1}s_nc_{n-1})^2.
    \end{aligned}
\end{equation}
The first two terms correspond to the renormalization of the photon-loss rate. The third term is a contribution to the relaxation rate mediated by the loss of photons, \textit{i.e.} Purcell effect \cite{Purcell,Blais2021} (Fig.~\ref{rates_renorm}(b)). The last one is a small contribution to the excitation rate (Fig.~\ref{rates_renorm}(e')) that we call ``cavity-relaxation-induced excitation'' since it involves the annihilation operator $a$.
Taking into account the relevant frequencies, one gets:
\begin{equation}
\label{EQ:down_a}
    \begin{aligned}
        \Gamma^{a}_\downarrow(n) &= \kappa^{0}[\omega_q]  (\sqrt{n}c_{n+1}s_{n}-\sqrt{n+1}s_{n+1}c_{n})^2,\\
\end{aligned}
\end{equation}
\begin{equation}
    \begin{aligned}
    \label{EQ:up_a}
\Gamma^{a}_\uparrow(n) &= \kappa^{0}[-\omega_q+2\omega_r] (\sqrt{n}c_ns_{n-1}-\sqrt{n-1}s_nc_{n-1})^2.   
    \end{aligned}
\end{equation}
The notation $\kappa^{0}[\omega_q]$ instead of $\kappa$ stresses the fact that photon relaxation is probing the environment at $\omega_q$ \cite{Blais2021}. 
When $n \ll n_{\rm{crit}}$ and $n_{\rm{crit}} \gg 1$,
\begin{equation}
\Gamma^{a}_\downarrow(n)\approx \Gamma_\kappa \left(1-\frac{3n}{2n_{\rm{crit}}}\right),
\end{equation}
where $\Gamma_\kappa=\kappa /4 n_{\rm{crit}}=\kappa \left(\frac{g}{\Delta}\right)^2$ is the Purcell rate, as derived in Ref.~\cite{Sete}. 

The rates of the inverse processes associated to the $a^{\dagger}$ operator, ``Inverse-Purcell excitation'' and ``cavity-excitation-induced relaxation'' (Fig.~\ref{rates_renorm}(b') and (e)), read:
\begin{equation}
\begin{aligned}
    \label{EQ:up_adag}
    \Gamma^{a^{\dagger}}_\uparrow(n) &= \kappa^{0}[-\omega_q]  (\sqrt{n}c_{n+1}s_{n}-\sqrt{n+1}s_{n+1}c_{n})^2;\\
\end{aligned}
\end{equation}
\begin{equation}
    \begin{aligned}
    \label{EQ:down_adag}
    \Gamma^{a^{\dagger}}_\downarrow(n) &= \kappa^{0}[\omega_q-2\omega_r] (\sqrt{n}c_ns_{n-1}-\sqrt{n-1}s_nc_{n-1})^2.
\end{aligned}
\end{equation}

\subsection{Effect of $\sigma_z$: dressed dephasing}
We now consider the action of $\sigma_z$, associated to fluctuations in the transition energy that produce dephasing of the undressed qubit. These contributions were named ``dressed dephasing'' in Ref.~\cite{Boissonneault}. The non-zero matrix elements of the dressed qubit are:
\begin{equation}
    \begin{aligned}
        |\overline{\bra{g,n}}\sigma_z \overline{\ket{g,n}}|^2 &=  (c_n^2-s_n^2)^2,\\
        |\overline{\bra{e,n}}\sigma_z \overline{\ket{e,n}}|^2 &=  (c_{n+1}^2-s_{n+1}^2)^2,\\
        |\overline{\bra{e,n-1}}\sigma_z \overline{\ket{g,n}}|^2 &=  4c_{n}^2s_{n}^2 \\
        |\overline{\bra{g,n+1}}\sigma_z \overline{\ket{e,n}}|^2 &=  4c_{n+1}^2s_{n+1}^2.
    \end{aligned}
\end{equation}
The first two describe the renormalization of the dephasing rate $\Gamma_\phi(n)$.  The third and fourth ones, which we call ``dephasing-induced excitation'' and ``dephasing-induced relaxation'', are contributions to the excitation and relaxation rates $\Gamma^{\sigma_z}_\uparrow(n)$ ($\Gamma^{\sigma_z}_\downarrow(n)$) (see Fig.~\ref{rates_renorm}(c)). One obtains:
\begin{equation}
\label{EQ:Gamma^sigmaz}
    \begin{aligned}
        \Gamma^g_\phi(n) &= \Gamma^{0}_{\phi}[0] R_{-}(n),\\
\Gamma^e_\phi(n) &= \Gamma^{0}_{\phi}[0] R_{-}(n+1),
\end{aligned}
\end{equation}
\begin{equation}
    \begin{aligned}
    \label{EQ:up_sigmaz}
\Gamma^{\sigma_z}_\uparrow(n)&= \Gamma^{0}_{\phi}[-\omega_q+\omega_r] R_{cs}(n),
\end{aligned}
\end{equation}
\begin{equation}
    \begin{aligned}
    \label{EQ:down_sigmaz}
\Gamma^{\sigma_z}_\downarrow(n)&= \Gamma^{0}_{\phi}[\omega_q-\omega_r] R_{cs}(n+1). \\
    \end{aligned}
\end{equation}

In the dispersive limit, 
\begin{equation}
    \begin{aligned}
        \Gamma_\phi(n) &\approx& \Gamma^0_\phi[0] \left(1-\frac{n}{n_{\rm{crit}}}\right),\\
\Gamma^{\sigma_z}_\uparrow(n) &\approx& \Gamma^0_\phi[-\omega_q+\omega_r] \frac{n}{n_{\rm{crit}}},\\
\Gamma^{\sigma_z}_\downarrow(n) &\approx& \Gamma^0_\phi[\omega_q-\omega_r] \frac{n}{n_{\rm{crit}}},
    \end{aligned}
\end{equation}
The first equation is used to calibrate the photon number in experiments with long coherence time qubits \cite{Gambetta}. The two others, as discussed by Boissoneault {\it et al.} \cite{Boissonneault,Boissonneault2009,Boissonneault2010}, explain how the qubit readout fidelity is affected when measuring at large power.

\section{Continuous measurements analysis}
\label{app:timetrace}
\setcounter{figure}{0}

\begin{figure}[t!]
\includegraphics[width=1\columnwidth]{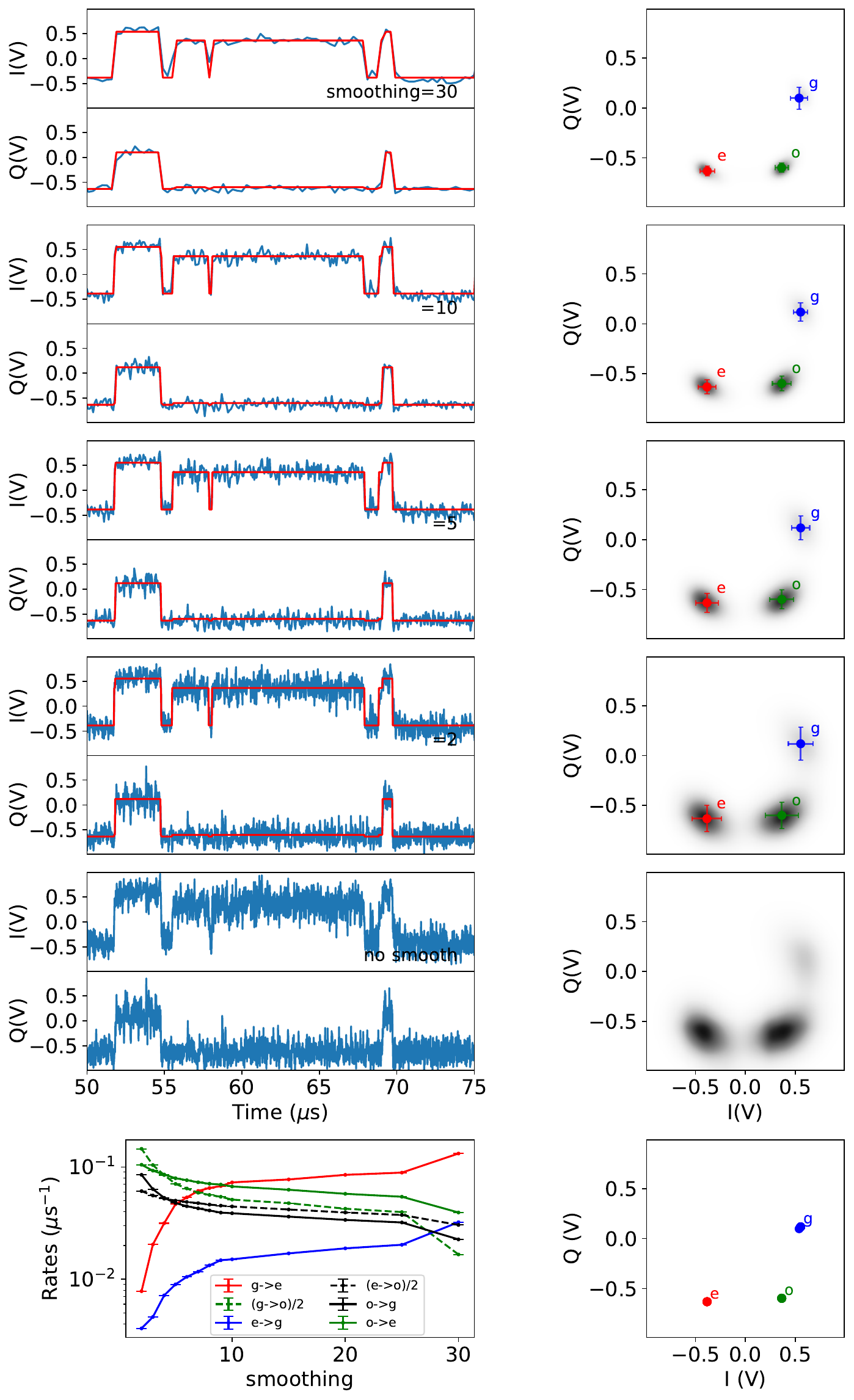}
\caption{\textbf{Effect of smoothing of the time traces, for strong measurement power.} The results of the HMM analysis are shown for decreasing smoothing factors (from top to bottom), for the contact at $f_A=7.5~$GHz, at $\overline{n}_g=163$ and $\overline{n}_o=$226: red lines indicate the most probable state at each time. Left panels shown the same excerpt of a trace ($I$ and $Q$ quadratures) with the various smoothing; right panels show the corresponding histograms in the $(I,Q)$ plane (grey scale), with the position of the clouds as found from the HMM analysis. Error bars indicate the size of the clouds. Bottom panels show on the left the evolution of the rates with smoothing; on the right the evolution of the position of the clouds.}
\label{smoothing}
\end{figure}
\begin{figure}[t!]
\includegraphics[width=1\columnwidth]{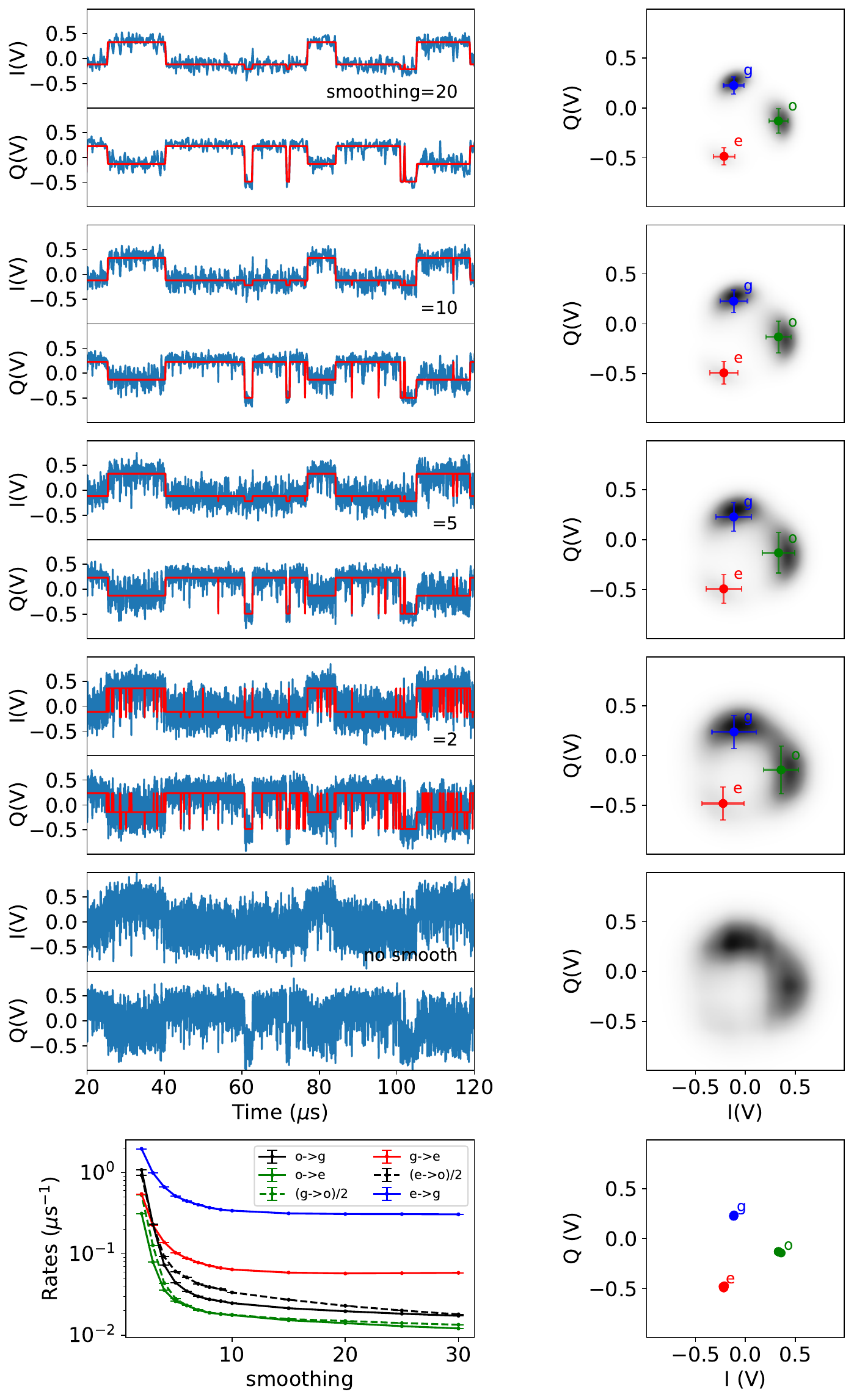}
\caption{Effect of the smoothing of the time traces, for weak measurement power: same parameters as Fig.~\ref{smoothing}, except that  $\overline{n}_g=10$ and $\overline{n}_o=$23.}
\label{smoothing2}
\end{figure}

\begin{figure*}[t!]
\includegraphics[width=2.1\columnwidth]{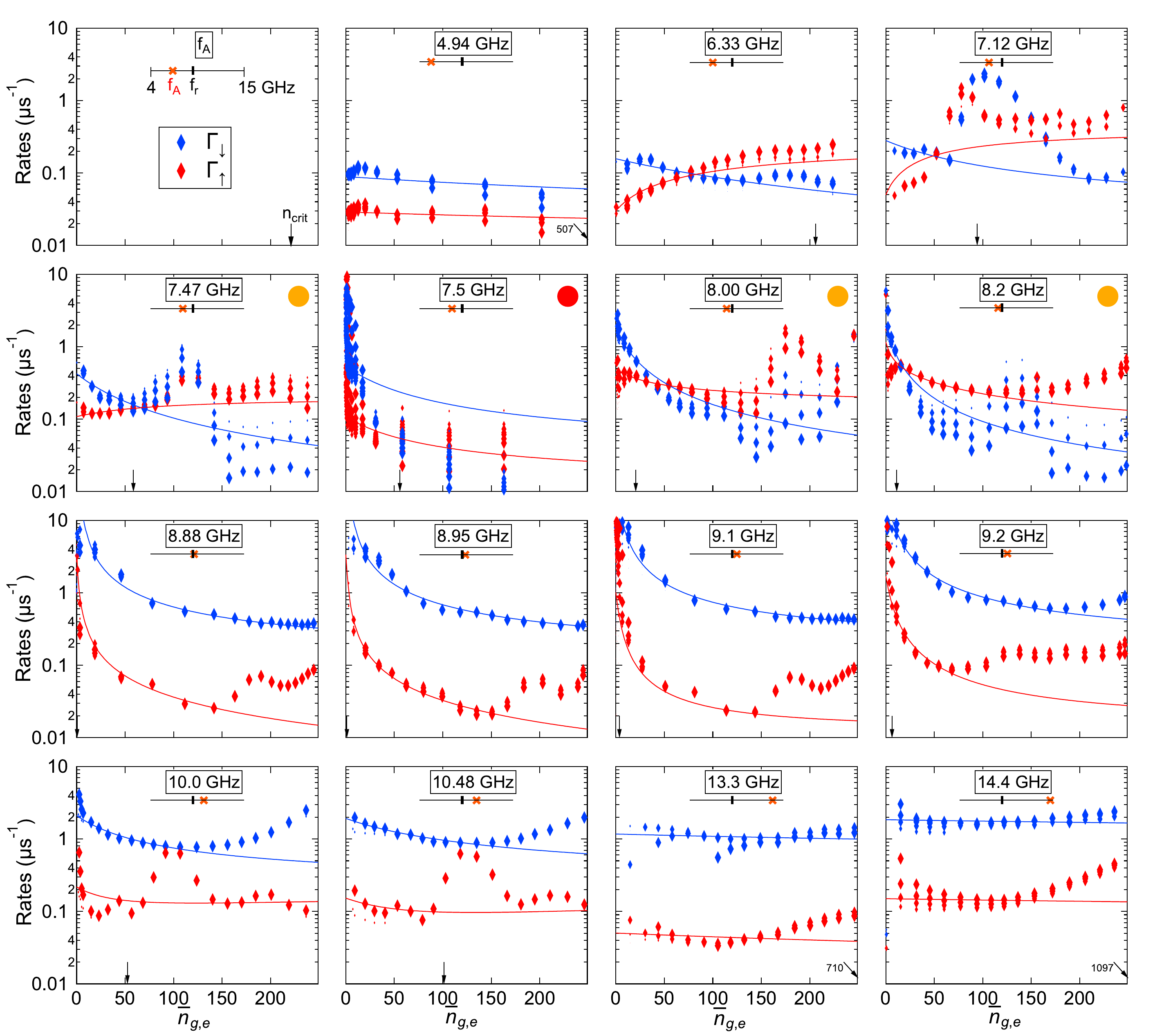}
\caption{\textbf{Relaxation and excitation transition rates as a function of photon number} for different contacts ($f_A$ indicated on each panel from $f_A=4.94$~GHz to 14.4~GHz, with symbolic representation of the relative position of $f_A$ (red cross) relatively to $f_r$ (black tick) on a segment representing the interval 4--15~GHz). Arrows on the x-axis indicate the value of $n_{\rm crit}$. Rates obtained from the analysis of time traces with less filtering are represented with bigger symbols (see text). Orange and red disks signal results that depend significantly on filtering, and hence are less significant.
Continuous lines correspond to calculated dependencies using the theoretical expressions in Eq.~(\ref{eq_fit_rates}) with prefactors shown in Fig.~\ref{ratesfreqfitcoeff}.}
\label{powerdepeven-complete}
\end{figure*}

\begin{figure*}
\includegraphics[width=2.1\columnwidth]{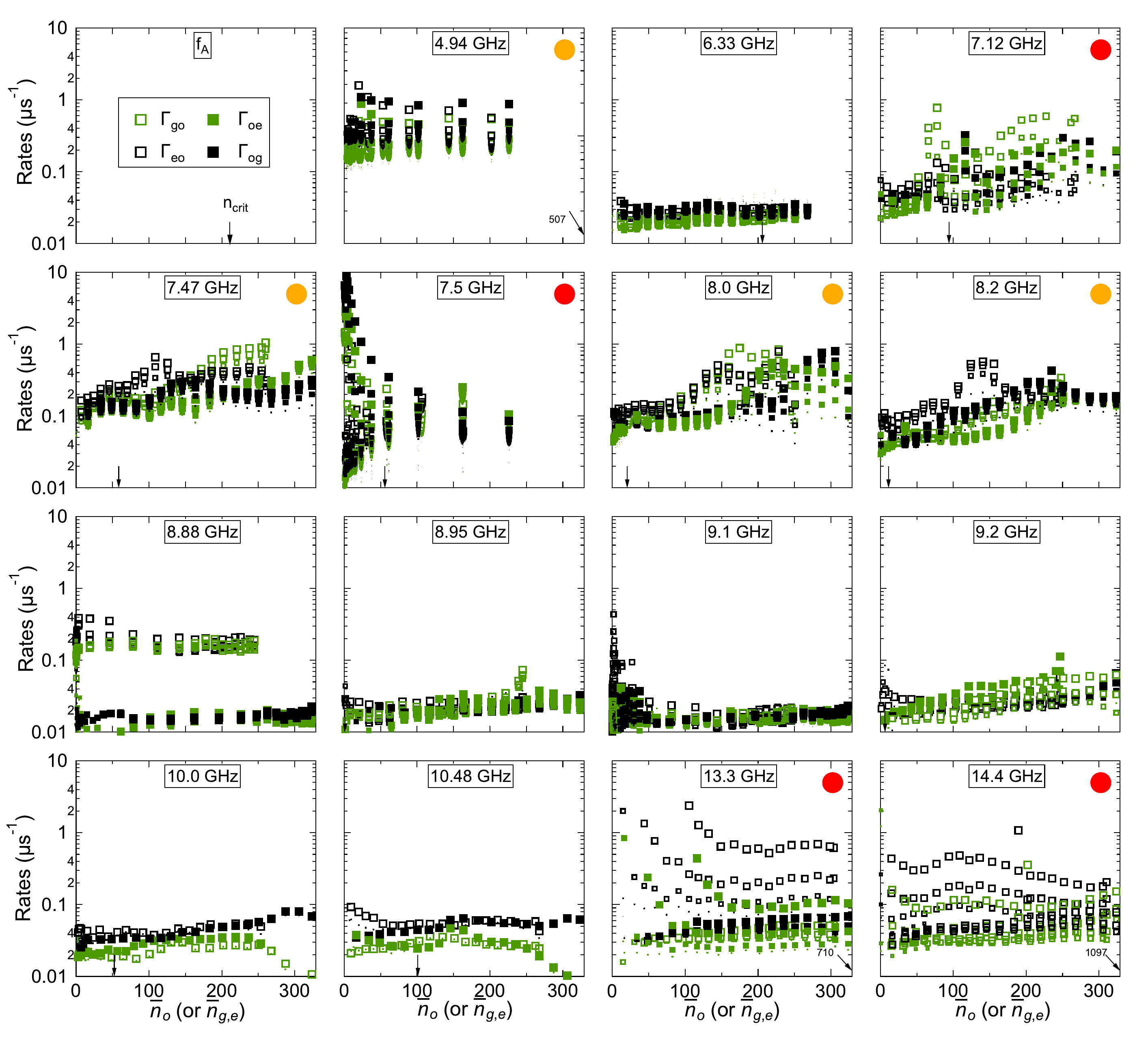}
\caption{\textbf{Parity jump transition rates as a function of photon number} for different contacts ($f_A$ indicated on each panel from $f_A=4.94$~GHz top-left to 14.4~GHz right-bottom). Rates obtained from the analysis of time traces with less filtering are represented with bigger symbols (see text). Orange and red disks signal results that depend significantly on filtering, and hence are less significant.}
\label{powerdepodd-complete}
\end{figure*}

The analysis of continuous measurements $I(t), Q(t)$ is delicate due to the presence of noise, in particular when the measurement tone amplitude is small and the values of $(I,Q)$ corresponding to the different states are too close. A way around is to average the data, and replace a series of $N_{\rm av}$ successive measurements by their mean value (``box averaging''), hence reducing the noise by a factor $\sqrt{N_{\rm av}}$. However, this has the drawback of filtering out fast transitions. Tests performed on computer-generated traces showed that, to determine the rates, the smaller $N_{\rm av}$ the better, as long as the signal to noise allows the determination. Another important parameter is the sampling rate, and its comparison with the signal filtering. In our case, the signal was filtered with a 60~MHz low-pass filter, so that, according to Shannon criterium, the sampling rate needs to be at least 120~MHz, \textit{i.e.} 8~ns/point. We used 10~ns/point, which is almost optimal. In practice, we compared the rates obtained by the HMM analysis when reducing $N_{\rm av}$ from 30 to 2. As exemplified in Figs.~\ref{smoothing} and \ref{smoothing2}, it is sometimes found that the rates extracted from the analysis show a smooth evolution when reducing the smoothing factor, then start diverging at low $N_{\rm av}$ when the signal-to-noise ratio reduces. However, in most cases, the position of the centers of the clouds as determined by the HMM is almost independent of $N_{\rm av}$. When it was not the case, the results were discarded. We show in Figs.~\ref{powerdepeven-complete} and \ref{powerdepodd-complete} the results obtained at increasing values of $N_{\rm av}$ with symbols of decreasing size, to illustrate the sensitivity of the rates to filtering. Large-size points rates correspond to a soft filtering (2 to 5 points), while smaller points correspond to 10- to 30-points averaging. For the data sets repeated in Figs.~\ref{powerdepeven} and \ref{powerdepodd}, the symbols of different sizes overlap, indicating that the rates obtained are independent on filtering. But for others, it is not the case, indicating that the rates determination is less or not reliable. We have indicated this with orange or red warning disks at the top right of the corresponding panels in Figs.~\ref{powerdepeven-complete} and \ref{powerdepodd-complete}. 

\section{Results of the ``fits''}
\setcounter{figure}{0}
\label{app:fit}
We present in Fig.~\ref{ratesfreqfitcoeff} the prefactors used for the theory curves plotted in Fig.~\ref{powerdepeven}. They were obtained by a manual adjustment to account at best for the overall dependence of the excitation and relaxation rates, with more weight given to the low photon number points. Since $\Gamma_\downarrow$ is found to decay with $\overline{n}$ at low $\overline{n}$, we essentially tried to reproduce the relaxation by a combination of dressed and Purcell relaxation terms. Symmetrically, excitation rates were mainly accounted for by a combination of dressed and inverse-Purcell excitation. 
To improve the agreement of the excitation rates at large $\overline{n}$, we included contributions increasing with $\overline{n}$, which can be done in different ways. For cases when $f_q<f_r$ the dephasing-induced excitation accounted best for the $\overline{n}$ dependence. In some cases we have also included cross-contributions (relaxation- and cavity-relaxation-induced excitation).

\begin{figure*}[t!]
\includegraphics[width=2\columnwidth]{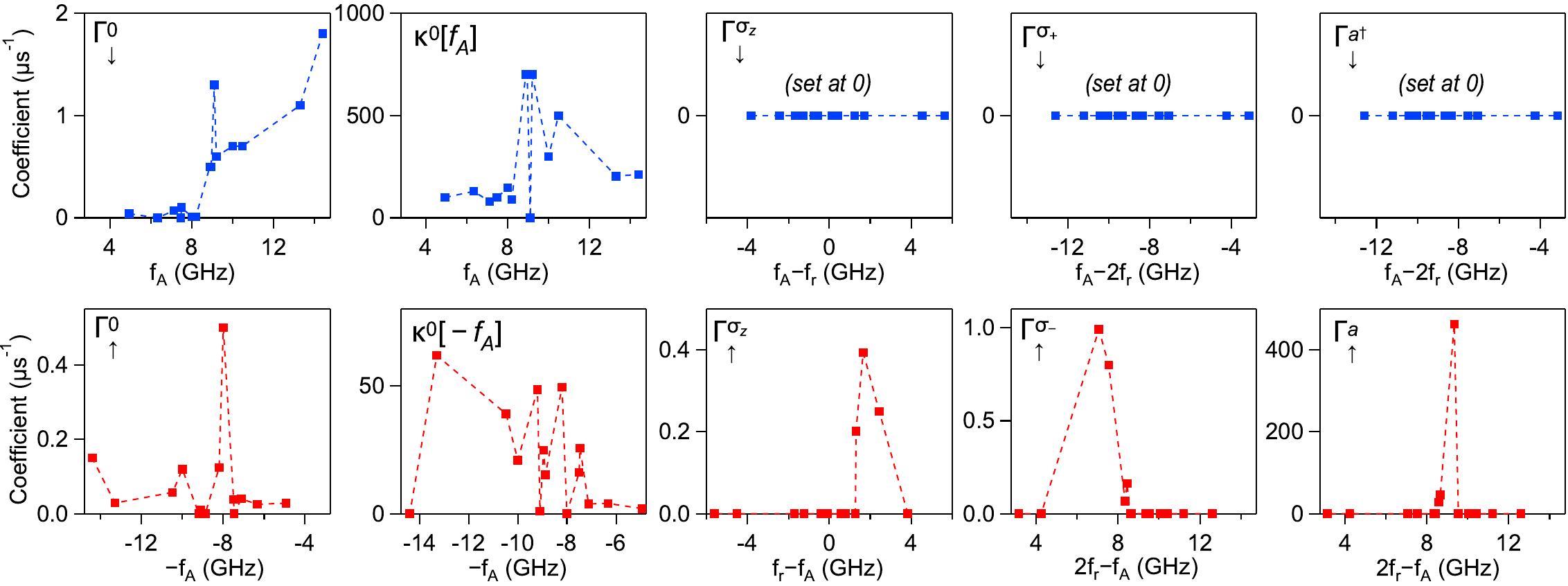}
\caption{Prefactors of the different contributions to the dressed dynamics (see Fig.~\ref{rates_renorm}) obtained from comparison with the data in Fig.~\ref{powerdepeven} as a function of the relevant frequencies. Top (bottom) panel:  dressed, Purcell (inverse Purcell), dephasing-induced, excitation- (relaxation-) induced and cavity-excitation-induced relaxation (cavity-relaxation-induced excitation). (The disposition of the panels corresponds to that in Fig.~\ref{rates_renorm}). }
\label{ratesfreqfitcoeff}
\end{figure*}

\section{Numerical simulations}
\label{app:ddrop}
\setcounter{figure}{0}
In order to check the validity of the analytical results for the power-dependent rates and to compare the DDROP experimental results with theory, we performed extensive
numerical simulations using the Python Qutip package \cite{QuTip}. We present in the following the Hamiltonian and the collapse operators, and how the cavity and qubit drives were treated. We introduce all this progressively and discuss the expected behavior.
\subsection*{Hamiltonian}
The Hamiltonian of the cavity+Andreev qubit system is 
\begin{equation}
H_0=H_c+H_A+H_g,
\end{equation}
where $ H_c=\hbar\omega_r\left(a^\dagger a+\frac{1}{2}\right)$, is the Hamiltonian of the cavity with resonance frequency $\omega_r/2\pi$,
$H_A=\frac{1}{2}\hbar \omega_A(\varphi,\tau)\sigma_z$,
is the Hamiltonian of the Andreev qubit in the Andreev basis $\{\ket{g},\ket{e}\}$ with $\hbar \omega_A(\varphi,\tau)=2\Delta_s\sqrt{1-\tau\sin^2{(\varphi/2)}}$ ($\Delta_s$ is the superconducting gap and $\tau$ the electronic transmissio   ²n coefficient in the channel and $\varphi$ the phase). Finally,
\begin{equation}
H_g=\varphi_{\text{zpf}}H'_A(a^{} + a^{\dagger}) + \varphi_{\text{zpf}}^2(H''_A)_z \left(a^{\dagger}a^{}+\frac{1}{2}\right)
\end{equation}
describes the coupling between the Andreev qubit and the cavity up to order $\varphi_{\text{zpf}}^2$ \cite{Park2020}. Here, the derivatives are with respect to $\varphi$ and the zero point phase fluctuations $\varphi_{\text{zpf}}$ are those imposed by the cavity zero point current fluctuations through the mutual inductance coupling to the superconducting loop (see Fig.~\ref{setup}). The operators are
\begin{equation}
    \begin{aligned}
        H'_A&=\varphi_0 I_A(\varphi,\tau)\left[\sigma_z+\sqrt{1-\tau}\tan{(\varphi/2)}\sigma_x\right],\\
        H''_A&=\varphi_0 I_A(\varphi,\tau)\left[\frac{\tau+(2-\tau)\cos\varphi}{2\sin\varphi}\sigma_z+\sqrt{1-\tau}\sigma_y\right],
    \end{aligned}
\end{equation}
where $I_A(\varphi,\tau)=\frac{\Delta_s}{4\varphi_0}\frac{\tau \sin\varphi}{\sqrt{1-\tau\sin^2(\varphi/2)}}$, and $\varphi_0=\hbar/2e$ the reduced flux quantum. In the experiments reported here,
$\varphi=\pi$ and $\tau\approx1$, so that
\begin{equation}
    \begin{aligned}
H'_A(\pi)&\approx\frac{\Delta_s\tau}{2} \sigma_x,\\
H''_A(\pi)&\approx -\frac{\Delta_s}{4}\tau\sqrt{1-\tau}\sigma_z,
    \end{aligned}
\end{equation}
and the diagonal contribution of $H''_A$ can be neglected for high-transmission. Then
\begin{equation}
    \begin{aligned}
H_0=\hbar\omega_r a^{\dagger} a^{} +\frac{\hbar \omega_A}{2}\sigma_z+\hbar g \sigma_x(a^{} + a^{\dagger}),
\label{Eq:H0}
    \end{aligned}
\end{equation}
with $\hbar g=\varphi_{\text{zpf}}\frac{\Delta_s\tau}{2}$.
\\
\paragraph{Rotating wave approximation---}
The coupling term
\begin{equation}
\hbar g \sigma_x(a^{} + a^{\dagger})= \hbar g (\sigma_+ + \sigma_-)(a^{} + a^{\dagger}),
\end{equation}
reads, in the interaction picture:
\begin{equation}
    \begin{aligned}
        &\hbar g \left[\sigma_+a^{} e^{i(\omega_A-\omega_r)t} + \sigma_-a^{\dagger}e^{-i(\omega_A-\omega_r)t}\right] \\
        &+ \hbar g \left[ \sigma_+a^{\dagger} e^{i(\omega_A+\omega_r)t} + \sigma_-a^{}e^{-i(\omega_A+\omega_r)t}\right],
    \end{aligned}
\end{equation}
Assuming $|\omega_A-\omega_r|\ll \omega_A+\omega_r$ (which does not hold for all the experimental data reported here), only the part that changes slowly is kept. In this case, 
\begin{equation}
H_{0,\text{RWA}}=\hbar\omega_r a^{\dagger} a^{} +\frac{\hbar \omega_A}{2}\sigma_z+\hbar g (\sigma_+ a^{} + \sigma_-a^{\dagger})
\label{Eq:H0RWA}
\end{equation}
takes the form of the Jaynes-Cummings Hamiltonian. The total number of excitations is $\hat{N}_{t}=a^{\dagger} a^{} + \sigma_+\sigma_- = a^{\dagger} a^{} + \frac{1}{2}\left(1+\sigma_z\right)$ and
\begin{equation}
[H_{0,\text{RWA}},\hat{N}_{t}]=\hbar g [(\sigma_+ a^{} + \sigma_-a^{\dagger}),\hat{N}_{t}]=0,
\end{equation}
which means that the Hamiltonian in RWA preserves the number of excitations. In the coupled base $\{|e,n-1\rangle,|g,n\rangle\}$, and shifting the energy zero reference,
\begin{equation}
    \begin{aligned}
h^n_{0,\text{RWA}}=\hbar\omega_r  n +\left(\begin{array}{cc}\hbar(\omega_A -\omega_r) & \hbar g \sqrt{n}\\
\hbar g \sqrt{n} & 0\end{array}\right),
    \end{aligned}
\end{equation}
which after diagonalization gives the eigenenergies $E_{\pm}(n)=n \hbar\omega_r + \frac{\hbar\Delta}{2} \pm \frac{\hbar}{2}\sqrt{\Delta^2+4g^2n}$ and the eigenvectors:
\begin{equation}
    \begin{aligned}
\overline{|e,n-1\rangle}&= \left(\begin{array}{c}c_n\\ -s_n\end{array}\right) ;\\
\overline{|g,n\rangle}&= \left(\begin{array}{c}s_n\\ c_n\end{array}\right),
    \end{aligned}
\end{equation}
with $c_n$ and $s_n$ given in the main text, see Eq.~(\ref{dressedstates}).
\\
\paragraph{Dispersive Limit---} 
When $g\ll|\Delta|$ the coupling can be treated as a perturbation. This dispersive limit gives
\begin{equation}
H_{0,\text{disp}}= \hbar\left(\omega_r + \chi_0\sigma_z\right) a^{\dagger}a^{} +\frac{\hbar }{2}\left(\omega_A+\chi_0\right)\sigma_z.
\end{equation}
As discussed in the main text, the dispersive approximation predicts an $n$-independent cavity pull $\chi_0=\frac{g^2}{\Delta}$. Note that, if 
$n \ll n_{\rm crit},$ $\theta_n\approx\left(\frac{g\sqrt{n}}{\Delta}\right)$ and the eigenenergies  read
\begin{equation}
    \begin{aligned}
E_{+}(n) &\approx n\hbar (\omega_r+\chi_0) + \frac{\hbar}{2}(\omega_A+\chi_0),\\
E_{-}(n) &\approx n\hbar (\omega_r-\chi_0)-\frac{\hbar}{2}(\omega_A+\chi_0),
    \end{aligned}
\end{equation}
so the cavity remains harmonic, with a qubit-state-dependent frequency. 
The qubit energy $\hbar\omega_A + (2n+1)\hbar  \chi_0$ changes with $n$ (Stark shift).

By diagonalizing the Hamiltonian in the RWA we verified that the cavity pull depends on $n$, contrary to the dispersive limit assumption. The form $\chi$ changes with $n$ can be found to be well described by $\chi(n)=\chi_0/\sqrt{1+(n/n_{\text{c}})}$, where $n_{\text{c}}$ tell us how fast the non-linearity is reached. It is better to think on the cavity as a non-linear object whose resonance frequency depends on the amplitude of the field inside.

\subsection*{Uncoupled case: Driving and environment}
We consider first $g=0$.
\\
\paragraph{Cavity drive---}
In the experiments a measurement tone at $\omega_0\sim \omega_r$ populates the cavity with photons. The corresponding cavity drive (cd) Hamiltonian reads:
\begin{equation}
H_{cd}=\hbar A_0 \cos(\omega_{0}t) \left(a^{} + a^{\dagger}\right),
\end{equation}
where $A_0$ is the amplitude of the cavity drive. We introduce $U_{\text{RF}}=e^{i \omega_0 a^{\dagger}a^{} t}$ to transform into the rotating frame where $\tilde{H}=U^{}H_{0}U^{\dagger}-i\hbar U^{}\dot{U}^{\dagger} $ gives, neglecting counter-rotating terms, 
\begin{equation}
\tilde{H}_{\text{c+cd}}=\hbar \delta a^{\dagger} a^{} + \frac{\hbar A_0}{2} \left(a^{} + a^{\dagger}\right),
\end{equation}
where $\delta=\omega_r-\omega_0$.
\\
\paragraph{Photon Loss---}
The coupling of the cavity to a bath
translates into a finite lifetime for photons. The intrinsic photon loss rate being $\kappa$, the master equation reads
\begin{equation}
\frac{d a^{}}{dt}= \frac{i}{\hbar} [\tilde{H}_r,a^{}] 
+ \frac{\kappa}{2} {\cal L}[a^{\dagger}] a^{},
\end{equation}
where
\begin{equation}
{\cal L}[a^{\dagger}]{\cal O}=2 a^{\dagger} {\cal O} a^{} -a^{\dagger} a^{} {\cal O} - {\cal O} a^{\dagger} a^{} 
\Rightarrow {\cal L}[a^{\dagger}]a= -a^{},
\end{equation}
and
\begin{equation}
\frac{d a^{}}{dt}= -\left(i \delta +\frac{\kappa}{2} \right) a^{} - i\frac{A_0}{2}.
\end{equation}
The resolution of this equation assuming $a(0)=0$ gives 
\begin{equation}
a(t)=\frac{A_0}{i \kappa-2 \delta}\left( 1-e^{(-(i \delta+\kappa/2)t)}\right)
\end{equation}
and 
\begin{equation}
a^{\dagger}(t)a(t)=\frac{A_0^2}{\kappa^2+4 \delta^2}\left( 1+e^{-\kappa t}-2 e^{-\kappa t/2} \cos(\delta t) \right).
\end{equation}
The stationary occupation of the cavity is
\begin{equation}
\bar{n}_{\text{ss}}= \frac{A_0^2}{\kappa^2+4 \delta^2}.
\end{equation}
At resonance, the amplitude of 
the cavity drive and the occupation of the cavity are related by
$A_0= \sqrt{\bar{n}_{\text{ss}}(0)}\kappa$.
We have verified this relation with numerical simulations.
\\
\paragraph{Qubit drive ---}
A tone at $\omega_1$ drives the Andreev qubit. Introducing $U_{\text{RFq}}=e^{i \frac{\omega_1}{2} \sigma_z t}$, the qubit drive
\begin{equation}
H_{qd}=\hbar A_1 \cos(\omega_{1}t) \left(\sigma_+ + \sigma_-\right)
\end{equation}
can be included in the rotating-frame (and neglecting counter rotating terms) as
\begin{equation}
\tilde{H}_{\text{q+qd}}=\frac{\hbar}{2}\beta+ \frac{\hbar A_1}{2} \left(\sigma_+ + \sigma_-\right),
\end{equation}
where $\beta=\omega_A-\omega_1$. In the absence of relaxation this drive induces Rabi oscillations on the qubit with a frequency $\sqrt{A^2_{1}+\beta^2}$.

\paragraph{Qubit relaxation and excitation---}
The Andreev qubit is subject to the effect of relaxation and excitation sources. The intrinsic relaxation (excitation) rate being $\Gamma_\downarrow^0$ ($\Gamma_{\uparrow}^0$), the master equation for the density matrix is
\begin{equation}
    \begin{aligned}
\frac{d \hat{\rho}_q}{dt}= &-\frac{i}{\hbar} [\tilde{H}_{q+qd},\hat{\rho}_q]\\ 
&+ \frac{\Gamma_\downarrow^0}{2} \left(2 \sigma_- \hat{\rho}_q \sigma_+ -\sigma_+ \sigma_-\hat{\rho}_q  - \hat{\rho}_q \sigma_+ \sigma_-\right)\\
&+ \frac{\Gamma_{\uparrow}^0}{2} \left(2 \sigma_+ \hat{\rho}_q \sigma_- -\sigma_- \sigma_+\hat{\rho}_q  - \hat{\rho}_q \sigma_- \sigma_+\right),
    \end{aligned}
\end{equation}
which gives the time evolution of the components of the density matrix:
\begin{equation}
    \begin{aligned}
\dot{\rho}_{ee}&=- i\frac{A_1}{2} \left(\rho_{ge} - \rho_{eg}\right) - \Gamma_\downarrow^0\rho_{ee}+\Gamma_{\uparrow}^0\rho_{gg},\\
\dot{\rho}_{eg}&=-\left(i\beta +\frac{\Gamma_\downarrow^0 +\Gamma_{\uparrow}^0}{2} \right)\rho_{eg} + i\frac{A_1}{2} \left( \rho_{ee}-\rho_{gg}\right),\\
\dot{\rho}_{ge}&=\left(i\beta - \frac{\Gamma_\downarrow^0 + \Gamma_{\uparrow}^0}{2}\right)\rho_{ge} - i\frac{A_1}{2} \left(\rho_{ee} - \rho_{gg}\right),\\
\dot{\rho}_{gg}&=i\frac{A_1}{2} \left(\rho_{ge} - \rho_{eg} \right) + \Gamma_\downarrow^0 \rho_{ee}-\Gamma_{\uparrow}^0\rho_{gg}.
    \end{aligned}
\end{equation}
(Since $\rho_{ee}+\rho_{gg}=1$, the last equation is redundant). In the stationary state,
\begin{equation}
(\Gamma_\downarrow^0+\Gamma_{\uparrow}^0)\rho_{ee}^{ss}=- i\frac{A_1}{2} \left(\rho_{ge}^{ss} - \rho_{eg}^{ss}\right) +\Gamma_{\uparrow}^0,
\end{equation}
and
\begin{equation}
    \begin{aligned}
\rho_{eg}^{ss}&= i\frac{A_1}{2\left(i\beta +\frac{\Gamma_\downarrow^0+\Gamma_{\uparrow}^0}{2} \right)} \left( \rho_{ee}^{ss}-\rho_{gg}^{ss}\right),\\
\rho_{ge}^{ss}&= -i\frac{A_1}{2\left(-i\beta + \frac{\Gamma_\downarrow^0+\Gamma_{\uparrow}^0}{2}\right)} \left(\rho_{ee}^{ss} - \rho_{gg}^{ss}\right),\\
\Rightarrow \rho_{ge}^{ss}-\rho_{eg}^{ss}&=\frac{-iA_1\left(\Gamma_\downarrow^0+\Gamma_{\uparrow}^0\right)}{\left(2\beta^2 + \frac{(\Gamma_\downarrow^0)^2}{2}\right)} \left(\rho_{ee}^{ss} - \rho_{gg}^{ss}\right)
    \end{aligned}
\end{equation}
so that
\begin{equation}
\rho_{ee}^{ss}= \frac{1}{2} - \left(\frac{1}{2}-\rho_{ee}^{th}\right) \frac{\left(2\beta^2 + \Gamma_1^2/2\right)}{A^2_{1} + 2\beta^2 + \Gamma_1^2/2}.
\end{equation}
Here $\rho_{ee}^{th}=\Gamma_{\uparrow}^0/(\Gamma_\downarrow^0+\Gamma_{\uparrow}^0)$ and $\Gamma_1=\Gamma_{\uparrow}^0+\Gamma_\downarrow^0=1/T_1$.\\

\paragraph{Qubit dephasing---}
Energy fluctuations lead to a dephasing rate $\Gamma_{\phi},$ and give rise to an additional term
\begin{equation}
\sim \Gamma_{\phi} \left(\sigma_z \hat{\rho}_q \sigma_z -\hat{\rho}_q\right),
\end{equation}
in the master equation, so that only the non-diagonal terms change
\begin{equation}
    \begin{aligned}
\dot{\rho}_{eg}&=-\left(i\beta +\frac{\Gamma_1 + 2\Gamma_{\phi}}{2} \right)\rho_{eg} + i\frac{A_1}{2} \left( \rho_{ee}-\rho_{gg}\right),\\
\dot{\rho}_{ge}&=\left(i\beta - \frac{\Gamma_1+2\Gamma_{\phi}}{2}\right)\rho_{ge} - i\frac{A_1}{2} \left(\rho_{ee} - \rho_{gg}\right).
    \end{aligned}
\end{equation}
In the stationary state,
\begin{equation}
\rho_{ge}^{ss}-\rho_{eg}^{ss}=\frac{-iA_1(\Gamma_1+2\Gamma_{\phi})}{\left(2\beta^2 + \frac{(\Gamma_1+2\Gamma_{\phi})^2}{2}\right)} \left(\rho_{ee}^{ss} - \rho_{gg}^{ss}\right)
\end{equation}
and therefore \cite{Ithier}
\begin{equation}
    \begin{aligned}
\rho_{ee}^{ss}&= \frac{1}{2} - \left(\frac{1}{2}-\rho_{ee}^{th}\right)\frac{1}{
\displaystyle 
1+\frac{A^2_{1}}{\Gamma_1 \Gamma_2 \left(1+(\beta / \Gamma_2)^2 \right)}},
    \end{aligned}
\label{rho_ee}
\end{equation}
where 
$\Gamma_2=\Gamma_1/2+\Gamma_{\phi}=1/T_2$,

This is the expression that describes the line-shape of qubit spectroscopy. We have verified that this fits the simulated time-evolution of a driven qubit in presence of relaxation, excitation and dephasing. 

\subsection*{Coupled cavity-qubit}
\paragraph{Hamiltonian---}
The Hamiltonian in presence of driving is
\begin{equation}
    \begin{aligned}
H&=\hbar\omega_r a^{\dagger} a^{} +\frac{\hbar \omega_A}{2}\sigma_z+\hbar g (a^{}\sigma_+ + a^{\dagger}\sigma_-)\\ &+\hbar A_0 \cos(\omega_{0}t) \left(a^{} + a^{\dagger}\right) +\hbar A_1 \cos(\omega_{1}t) \left(\sigma_+ + \sigma_-\right),
    \end{aligned}
\end{equation}
and the evolution has to be solved together with the operators of photon loss, relaxation, excitation and pure dephasing. In the rotating frame of both drives,
\begin{equation}
    \begin{aligned}
\tilde{H}=&\hbar\delta a^{\dagger} a^{} +\frac{\hbar \beta}{2}\sigma_z\\
&+\hbar g (\sigma_+e^{i\omega_1t}+\sigma_-e^{-i\omega_1t})(a^{}e^{-i\omega_0t} + a^{\dagger}e^{i\omega_0t})\\
&+\hbar \frac{A_0}{2} \left[ a^{} + a^{\dagger}+ a^{}e^{-i2\omega_0t} +  a^{\dagger}e^{i2\omega_0t}\right]\\
&+\hbar \frac{A_1}{2} \left[ \sigma_- + \sigma_+ + \sigma_-e^{-i2\omega_1t} +  \sigma_+e^{i2\omega_1t}\right],
    \end{aligned}
\end{equation}
which can be simplified neglecting the fast rotating terms
\begin{equation}
    \begin{aligned}
\frac{\tilde{H}_{\text{RWA}}}{\hbar}=&\delta a^{\dagger} a^{} +\frac{\beta}{2}\sigma_z + \frac{A_0}{2} \left( a^{} + a^{\dagger}\right) +\frac{A_1}{2} \left( \sigma_- + \sigma_+ \right)\\
&+ g \left(\sigma_+a^{}e^{i(\omega_1-\omega_0)t}+\sigma_-a^{\dagger}e^{-i(\omega_1-\omega_0)t}\right),
    \end{aligned}   
\end{equation}
which is time-dependent. It reproduces all the features discussed for $g=0$ and has advantage that for $g\neq 0$ the effect of non-linearity and rates renormalization are taken into account. The disadvantage is the computational cost of the simulations. In order to reach the steady state, the time-evolution has to be continued to very long times but in rather small steps. In additions, for very large photon number, the matrix size $\left(2\times N_{\text{Fock}}\right)^2$ increases significantly the computation time. To overcome this difficulty, we used a cluster to parallelize the calculation.

\begin{figure}[h!]
\includegraphics[width=0.8\columnwidth]{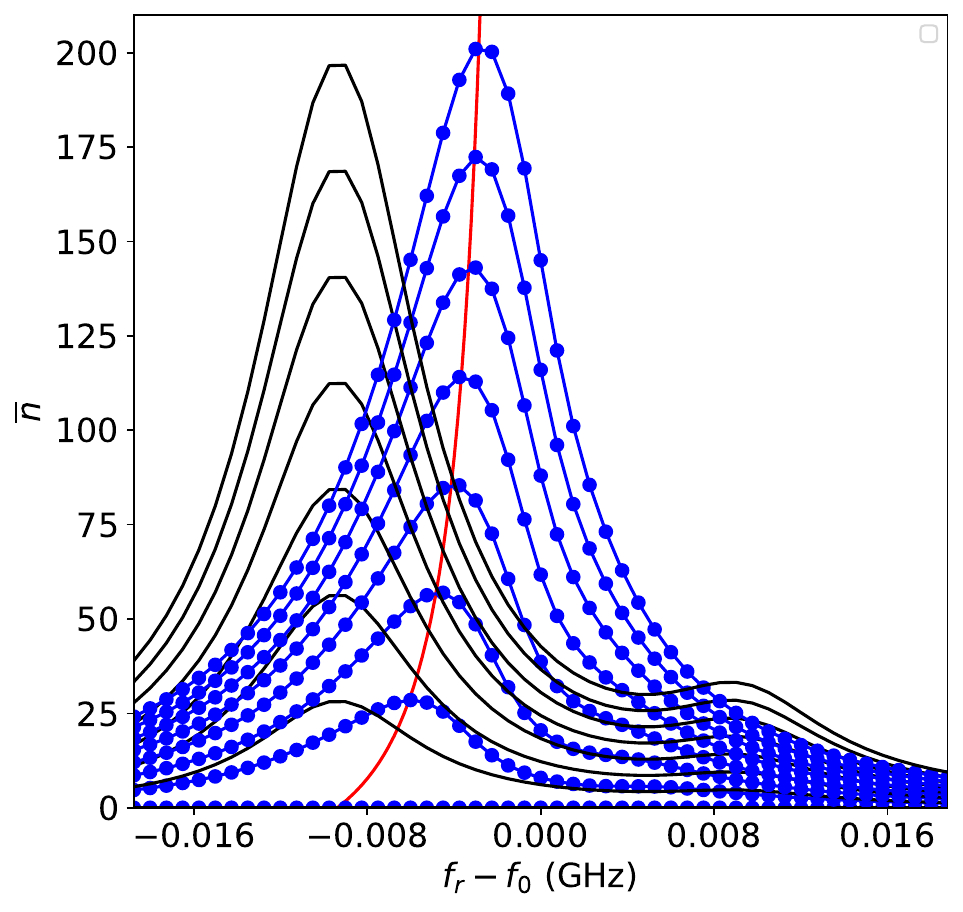}
\caption{Cavity mean occupation vs cavity-drive detuning $\delta=f_r-f_0$ for $g/2\pi=85$\ MHz, $f_q=8~$GHz at increasing drive power, in blue. In black lines the result in the dispersive limit, with a constant resonance shift $\chi_0=-0.009$~GHz. Red line shows $\chi(n)$.}
\label{app7}
\end{figure}

\paragraph{Driven cavity---}
We first checked the result of driving the cavity rendered non-linear by the coupling to the qubit. The mean steady-state occupation of the cavity in shown in Fig.~\ref{app7} as a function of the detuning $\delta$ for different drive amplitudes. 
The results of the simulation are compared with the dispersive limit where the cavity pull is constant, equal to $\chi_0$. The photon number $\left<n\right>$ increases when the drive frequency matches the qubit frequency in the ground state: $f_0-f_r=\chi_0$ There is a also a slight increase of $\left<n\right>$ at $f_0-f_r=-\chi_0$ due to the assumed thermal occupation of the excited state of the qubit $p_e=0.1$. With the simulation performed beyond the dispersive limit, one observes that the resonance shift reduces with the drive amplitude. The position of the maximum of the resonance follows $\chi(n)=\chi_0/\sqrt{1+n/n_{\rm crit}}$ (indicated with a red line), as expected. We have verified that the results of the simulation obey Eq.~(\ref{ndisperse}). 

\paragraph{Rates Renormalization I ---}
This is the most subtle point to take into account for the simulations. The master equation for the total density matrix $\hat{\rho}_{rq}$ is
\begin{equation}
    \begin{aligned}
\frac{d \hat{\rho}_{rq}}{dt}=& -\frac{i}{\hbar} [\tilde{H},\hat{\rho}_{rq}] 
+ \kappa {\cal D}[a^{}]\hat{\rho}_{rq} \\
+& \Gamma_\downarrow^0 {\cal D}[\sigma_-]\hat{\rho}_{rq} +  \Gamma_{\uparrow}^0 {\cal D}[\sigma_+]\hat{\rho}_{rq}+\Gamma_{\phi} {\cal D}[\sigma_z]\hat{\rho}_{rq},
    \end{aligned}
\end{equation}
where ${\cal D}[\hat{O}]\cdot =\hat{O}\cdot \hat{O}^{\dagger} -\frac{1}{2}\{\hat{O}^{\dagger} \hat{O}^{}, \cdot\}$ and this equation is valid for small coupling $g \ll \omega_r,\omega_A$ because it describes process in the bare base. As discussed in the main text, since the Hamiltonian mixes cavity and qubit states, there are ``new'' decay rates and rates renormalization.

\begin{figure}[h!]
\includegraphics[width=0.6\columnwidth]{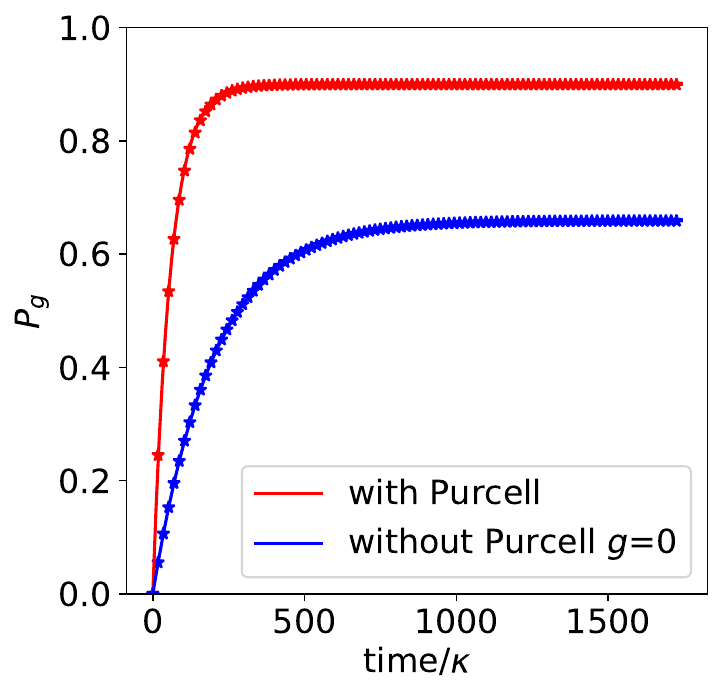}
\caption{Relaxation experiment: the qubit is initially in the excited state and evolves towards the thermal population in a time scale given by $T_1$. Red (blue) points correspond to the simulation for $g/2\pi=85$\ MHz ($g=0$), while lines are exponential fits (same parameters as in Fig.~\ref{app7}).}
\label{app8}
\end{figure}

This is particularly important to simulate a situation close to the experiments. 
Consider a qubit for which the lifetime and dephasing time are $T_1=1/\Gamma_1$ and $T_2$, with a thermal population $p_{th}$. When the qubit is not coupled to the cavity, the evolution is correctly simulated when one inputs: 
\begin{equation}
    \begin{aligned}
    \Gamma_{\uparrow}^0 &= p_{th} \Gamma_1 \\
    \Gamma_\downarrow^0 &= \Gamma_1-\Gamma_{\uparrow}^0.
    \end{aligned}
\end{equation}
This is no longer the case for the coupled system, as shown in Fig.~\ref{app8} with simulations of the coupled and uncoupled system using the same values for $\Gamma_\downarrow^0$ and $\Gamma_{\uparrow}^0$. The relaxation time and asymptotic population of the excited state are different.

This can be understood by considering the master equation written in the dispersive regime by Boissonneault, Gambetta and Blais \cite{Blais2021} (without any excitation rate):
\begin{equation}
    \begin{aligned}
\frac{d \hat{\rho}_{\text{disp}}}{dt}=& -\frac{i}{\hbar} [\tilde{H}_{\text{disp}},\hat{\rho}_{\text{disp}}] 
+ \left(\kappa + \kappa_{\Gamma_1}\right) {\cal D}[a^{}]\hat{\rho}_{\text{disp}} \\
&+ (\Gamma_\downarrow^0 + \Gamma_{\kappa}) {\cal D}[\sigma_-]\hat{\rho}_{\text{disp}} +\Gamma_{\phi} {\cal D}[\sigma_z]\hat{\rho}_{\text{disp}}\\
&+ \Gamma_{\Delta} {\cal D}[a^{\dagger}\sigma_-]\hat{\rho}_{\text{disp}}+ \Gamma_{\Delta} {\cal D}[a^{}\sigma_+]\hat{\rho}_{\text{disp}}, 
    \end{aligned}
\end{equation}
which shows that the Purcell rate $\Gamma_\kappa=\left(\frac{g}{\Delta}\right)^2\kappa$ directly adds to $\Gamma_\downarrow^0$. In addition, this equation introduces an increase of the cavity linewidth $\kappa_{\Gamma_1}=\left(\frac{g}{\Delta}\right)^2\Gamma_1$ and $\Gamma_\Delta=2\left(\frac{g}{\Delta}\right)^2\Gamma_\phi$ the dressed dephasing. Adding the excitation rate, one finally obtains $\Gamma_1=\Gamma_\downarrow^0+\Gamma_{\uparrow}^0+\Gamma_\kappa$ and $p_{th}=\Gamma_{\uparrow}^0/\Gamma_{1}$, which correctly describe the results of Fig.~\ref{app8}.

\paragraph{Rates Renormalization II ---}
As discussed in the main text, the rates are renormalized with the number of photons in the cavity. 
We followed the work of Sete, Korotkov and Gambetta \cite{Sete} to analyze this effect in details. We illustrate this analysis with the Purcell contribution:

\begin{itemize}
    \item fix the qubit frequency $f_q$, the cavity frequency $f_r$, the coupling $g$ and the photon loss rate $\kappa$.
    \item fix the frequency of the cavity-drive and the amplitude. Fix $\beta=\Delta+\delta$ to eliminate the explicit time-dependence. For example, for $\delta=0$, the Hamiltonian is
    \begin{equation}
        \frac{\tilde{H}_{\text{RWA}}}{\hbar}=\frac{\hbar \Delta}{2}\sigma_z + \frac{A_0}{2} \left( a^{} + a^{\dagger}\right) + g \left(\sigma_+a^{}+\sigma_-a^{\dagger}\right),
    \end{equation}
    \item Solve the equation $\overline{n}(\chi(\overline{n}))$ and build the initial state as a coherent state 
    \begin{equation}
    |\Psi_0\rangle=\sum_n P_{\overline{n}}(n) \overline{|e,n\rangle}
    \end{equation}
    \item compute the time-evolution and measure the expectation value of the projector $P_e=\sum_n\overline{|e,n\rangle} \overline{\langle e,n|}$ as a function of time. Fit the exponential decay and extract the rate.
    \item Repeat for a different drive amplitude.    
\end{itemize}

\begin{figure*}[t!]
\includegraphics[width=2\columnwidth]{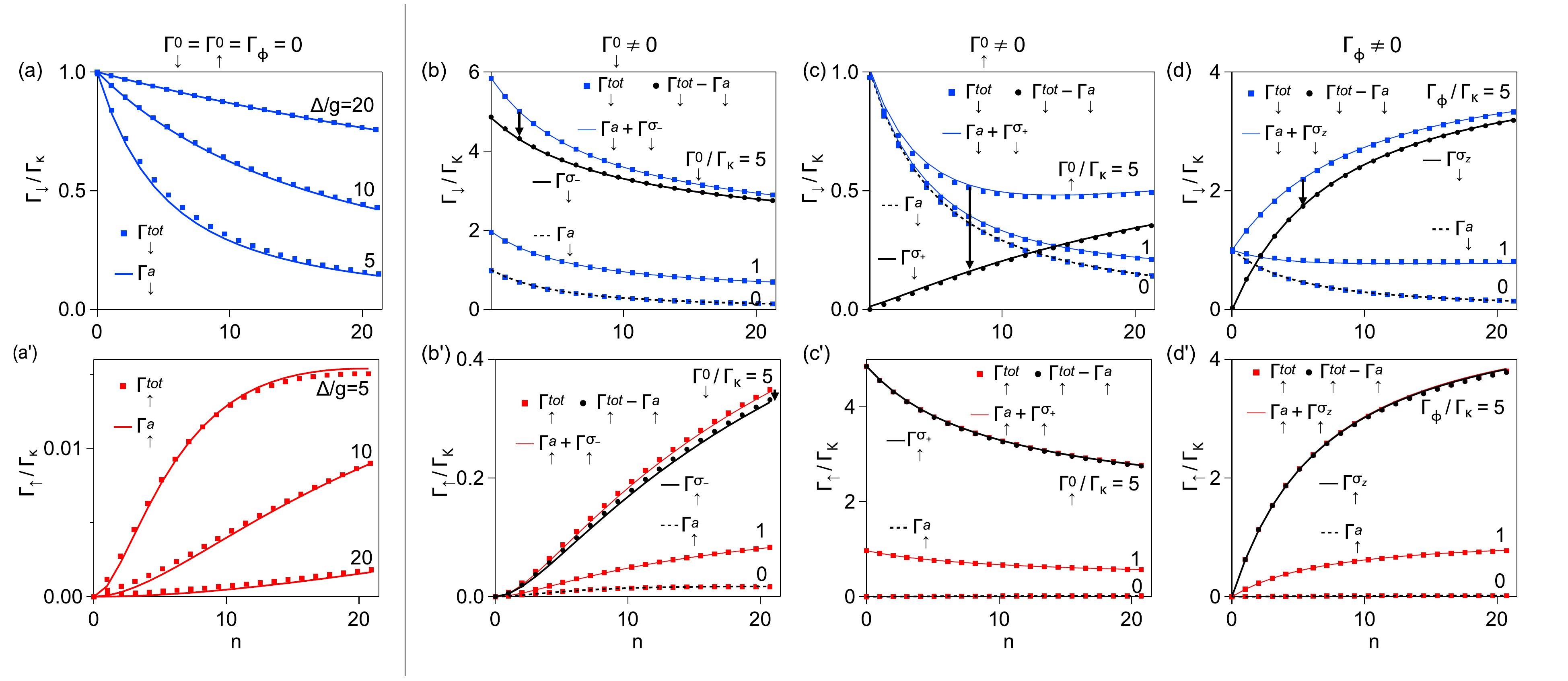}
\caption{Comparison between the rates renormalization obtained from numerical simulations of the time evolution (symbols), and the theoretical prediction (lines). (a,a') Renormalization of \textbf{Purcell relaxation and cavity-relaxation-induced excitation rates} (see Figs.~\ref{rates_renorm}(b) and (e')) for different values of $|\Delta|/g$, all other contributions being set to zero. The Purcell term is always present when one simulates the effect of the other terms. In the other panels, $\Delta/g=5$, and the corresponding curves $\Gamma_\downarrow^a$ and $\Gamma_{\uparrow}^{a^\dagger}$ are recalled with black dashed lines. Simulated rates are represented with blue and red squares. In each panel, we also show for the largest value of the parameter ($\Gamma_\downarrow^0$, $\Gamma_{\uparrow}^0$ or $\Gamma_{\phi}$ $=5\Gamma_{\kappa}$) the result of the simulation with the Purcell contribution subtracted (operation symbolized with a black arrow). (b,b') Renormalization of relaxation and excitation rates for $\Gamma_{\downarrow}^{0}/\Gamma_{\kappa}=$5, 1, and 0 (b) corresponds to \textbf{dressed relaxation} (see Fig.~\ref{rates_renorm}(a)) and (b') to \textbf{relaxation-induced excitation} (see Fig.~\ref{rates_renorm}(d')).
(c,c'). Renormalization of relaxation and excitation rates for different values of $\Gamma_{\uparrow}^{0}/\Gamma_{\kappa}$. (c) corresponds to \textbf{excitation-induced relaxation} (see Fig.~\ref{rates_renorm}(d)) and (c') to \textbf{dressed excitation} (see Fig.~\ref{rates_renorm}(a')). (d,d') Renormalization of relaxation and excitation rates for different values of $\Gamma_{\rm{\varphi}}/\Gamma_\kappa$. (d) corresponds to \textbf{dephasing-induced relaxation} (see Fig.~\ref{rates_renorm}(c)) and (d') to \textbf{dephasing-induced excitation} (see Fig.~\ref{rates_renorm}(c')). \textit{Parameters:} $\kappa/2\pi= f_r/950$, $g/2\pi=50$~MHz, $f_r=8.77$~GHz. We note that in this treatment of the time evolution for the open system the frequency dependence of the noise spectra are not taken into account.}
\label{app13}
\end{figure*}

In Figure \ref{app13} we show the numerical result of this procedure and the comparison with the theory of the renormalization discussed in the main part of the text. 
As soon as the qubit is coupled to the cavity, the relaxation and excitation rates have a Purcell contribution, which have their own dependence noted $\Gamma_\downarrow^a(n)$ on the photon number, as shown in (a,a'). It adds to the other processes, as shown in panels (b-d'). In (b,b'), one assumes a non-zero relaxation rate $\Gamma_\downarrow^0$. According to theory, this relaxation rate is dressed, a term that we have noted $\Gamma_\downarrow^{\sigma_-}(n)$. The simulation gives a total relaxation rate $\Gamma_\downarrow^a(n)+\Gamma_\downarrow^{\sigma_-}(n)$. Additionally, the non-zero $\Gamma_\downarrow^0$ leads to an excitation rate $\Gamma_{\uparrow}^{\sigma_-}(n),$ as shown in (b').
The main message is that the simulated time evolution gives a rate renormalization in agreement with the analytical expressions.
We have performed a similar verification of the renormalization of the relaxation and excitation rates and the contribution of the dressed dephasing.

\begin{figure*}[t!]
\includegraphics[width=2\columnwidth]{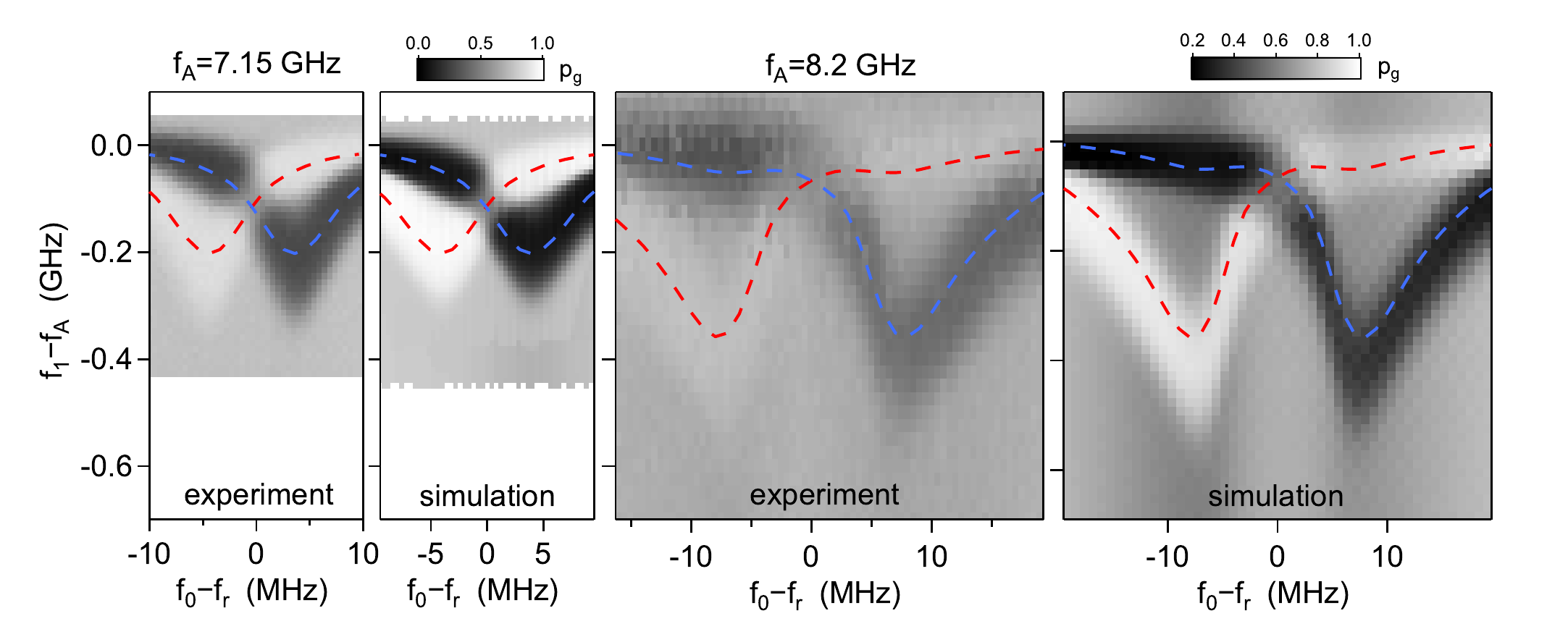}
\caption{Comparison of experimental data presented in Fig.~\ref{ddrop2} with DDROP simulations, for $f_A=7.15$~GHz and $f_A=8.2$~GHz. The parameters chosen for the simulation are $g/2\pi =85~$MHz, $\kappa = 58\mu$s$^{-1}$, $A_{1}/2\pi= 1.05$ MHz, and $\Gamma_{\downarrow}=0.116 ~\mu$s$^{-1}$ ($0.12~\mu$s$^{-1}$), $\Gamma_{\uparrow}=0.156~\mu$s$^{-1}$ ($0.05~\mu$s$^{-1}$), $\Gamma_{\phi}= 0.00937~\mu$s$^{-1}$ ($0.016~\mu$s$^{-1}$), for $f_A=8.2$~GHz ($f_A=7.15$~GHz) respectively, leading to $p_{th}=0.1$ (0.15), $T_1=0.640 \mu$s ($3 \mu$s) and $T_2=1.250~\mu$s ($5.45~\mu$s).} 
\label{app11}
\end{figure*}

\paragraph{Qubit drive in the presence of photons. DDROP protocol---}
\label{DDROP_app}
The simultaneous drive of the qubit and the cavity is the basis of the DDROP protocol as described in the main text. Since $\kappa$ is very large, we can imagine that the first that happens is that the cavity goes to a coherent state with a mean number of photons corresponding to the drive amplitude and the detuning. The qubit has a shifted frequency (a frequency distribution in fact weighted by the coherent state) and renormalized rates. As seen in the experimental data in Fig.~\ref{ddrop2} the center of the qubit frequency distribution follows $f_1=f_A+2 n_{g,e} \chi (\overline{n})$ according to Eqs.~(\ref{chiofn},\ref{n_vs_ain}) (dashed lines). 
Full simulations of the DDROP and their comparison with the experimental data are shown in Fig.~\ref{app11}. The positions of the resonances is well accounted for, but the values of the population $p_g$ of the ground state are different. Several factors explain this discrepancy. Firstly, the simulations ignore transition to $\ket{o}$. Secondly, the values of $\Gamma_\phi$ used in the simulation are much smaller than the measured ones, because the simulation does not take into account a frequency-dependent noise spectrum. Including the measured $\Gamma_\phi$, which is related to zero-frequency noise, would lead to very large dephasing-induced excitation and relaxation rate in the presence of photons, which is non-realistic because these terms depend on noise at frequencies $\pm (f_q-fr).$ The incorrect value of $\Gamma_\phi$ leads to an incorrect stationary value of $p_e$ (see Eq.~(\ref{rho_ee})). Thirdly, relaxation during the 1-$\mu$s-long measurement pulse is not taken into account.

~

\begin{acknowledgments}
Initial steps of this project were carried out by C. Janvier during his PhD thesis. We gratefully acknowledge B.~Huard for providing us with the JPC used in the experiments. We thank P. S\'enat and P.-F. Orfila for technical assistance. We acknowledge A. Levy Yeyati for constant theoretical support, A. Blais, G. Catelani, M. Devoret, A. Di Paolo, Yu. Nazarov and A. Petrescu for useful discussions. This work was supported by ANR contract JETS, and by FET-Open contract AndQC. L. Tosi was supported by the Marie Sk\l{}odowska-Curie individual fellowship grant 705467. 
\end{acknowledgments}

\end{document}